\documentclass[showpacs,preprintnumbers,amsmath,amssymb,twocolumn,nofootinbib]{revtex4}

\usepackage{graphicx}
\usepackage{dcolumn}
\usepackage{bm}
\usepackage{hyperref}
\usepackage{epsfig,slashed}

\newcommand{\nco}{\newcommand}
\nco{\beq}{\begin{equation}} \nco{\eeq}{\end{equation}}
\nco{\beqa}{\begin{eqnarray}} \nco{\eeqa}{\end{eqnarray}}
\nco{\lra}{\leftrightarrow}
\def\sfrac#1#2{{\textstyle{#1\over #2}}}

\nco{\sss}{\scriptscriptstyle} \nco{\dphi}{\varphi}
\nco{\lsim}{\mbox{\raisebox{-.6ex}{~$\stackrel{<}{\sim}$~}}}
\nco{\gsim}{\mbox{\raisebox{-.6ex}{~$\stackrel{>}{\sim}$~}}}
\nco{\etal}{\textit{et al.}}
\nco{\ud}{\mathrm{d}}

\begin{document}

\preprint{SACLAY--T10/062}
\preprint{CERN-PH-TH/2010-097}

\title{Can multistate dark matter annihilation explain\\ the high-energy cosmic
ray lepton anomalies?}

\author{Marco Cirelli$^{1,2}$, James M.\ Cline$^{1,3}$}

\affiliation{%
\centerline{$^{1}$CERN Theory Division, CERN, 
Case C01600, CH-1211 Gen\`eve, Switzerland}}
\affiliation{%
\centerline{$^{2}$Insitut de Physique Th\'eorique, CNRS 
URA 2306 \& CEA/Saclay, F-91191 Gif-sur-Yvette, France}}

\affiliation{%
\centerline{$^{3}$Physics Department, McGill University,
3600 University Street, Montr\'eal, Qu\'ebec, Canada H3A 2T8}
e-mail: marco.cirelli@cea.fr, jcline@physics.mcgill.ca }

\date{17 May 2010}

\begin{abstract} 

Multistate dark matter (DM) models with small mass splittings and
couplings to light hidden sector bosons have been proposed as an
explanation for the PAMELA/Fermi/H.E.S.S.\ high-energy lepton
excesses.  We investigate this proposal over a wide range of DM
density profiles, in the framework of concrete models with doublet or
triplet  dark matter and a hidden SU(2) gauge sector that mixes with
standard model hypercharge.   The gauge coupling is bounded from
below by the DM relic density, and the Sommerfeld enhancement factor
is explicitly computable for given values of the DM and gauge boson
masses $M$, $\mu$ and the (largest) dark matter mass splitting
$\delta M_{12}$.   Sommerfeld enhancement is stronger at the galactic
center than near the Sun because of the radial dependence of the DM
velocity profile, which strengthens the inverse Compton (IC) gamma
ray constraints relative to usual assumptions.  We find that the 
PAMELA/Fermi/H.E.S.S.\ lepton excesses are marginally compatible with
the model predictions, and with CMB and Fermi gamma ray constraints,
for $M\cong 800$ GeV, $\mu\lsim 200$ MeV, and a dark matter profile
with noncuspy Einasto parameters $\alpha\gsim 0.20$, $r_s \sim 30$
kpc. We also find that the annihilating DM must provide only  a
subdominant ($\lsim 0.4$) component of the total DM mass density,
since otherwise the boost factor due to Sommerfeld enhancement is too
large.

\end{abstract}

\maketitle

\section{Introduction}

Dark matter (DM) annihilating in the galaxy has been proposed as the
explanation for a number of current experimental anomalies.   The
Payload for Antimatter-Matter Exploration and Light-nuclei
Astrophysics (PAMELA) experiment has observed an excess in the
positron fraction in the energy range $10-100$ GeV
\cite{pamela}, and the Fermi Large Area Telescope (LAT) \cite{fermi}  
observes an excess in $e^++e^-$ at
energies up to 1 TeV, compared to the predictions of conventional
diffusive models. 
The High Energy Stereoscopic System (H.E.S.S.) \cite{hess}, too, observes a steepening in the spectrum around a few TeV which is in agreement with the Fermi-LAT excess. 
On the other hand, PAMELA does not observe excess
antiprotons.  Therefore  TeV-mass particles that can annihilate
preferentially into leptons are the preferred DM candidates for
explaining these anomalies \cite{CKRS,donato,cholis,berg,meade}.

Theories of leptophilic annihilating DM have come under increasing
pressure from a number of complementary constraints, due to the radio
and gamma ray (GR) emissions that should be produced by the annihilation process itself (`prompt' gamma rays) \cite{prompt1,prompt2,prompt3,prompt4} or by the high
energy $e^\pm$, via synchrotron radiation in the galactic magnetic
field or via inverse Compton scattering on cosmic microwave
background (CMB) photons and the galactic radiation field (starlight)
\cite{CP,cuoco,Regis:2009md,Cholis:2009gv,papucci,CPS,zaharijas,hutsi}.  Such radiation could also partially reionize the
universe after recombination, leading to distortions in the CMB
\cite{galli,SPF,hutsiCMB,CIP,kanzakiCMB}. 
Finally, relevant constraints are imposed by extragalactic gamma rays, produced by the annihilations in forming DM halos \cite{Profumo,Belikov:2009cx,hutsiCMB}. 
Many models of annihilating DM are seemingly ruled out by overproducing such radiation. 
In particular, inverse Compton gamma rays produced in the galaxy would contribute to the GR spectrum observed by Fermi-LAT.   

A precise assesment of the strength of Fermi inverse Compton (IC) constraints depends however upon the details of the assumed DM density profile $\rho(r)$. 
The strongest constraints come from close to the Galactic Center (GC); 
therefore they can be satisfied if $\rho(r)$ is not strongly peaked
as $r\to 0$, where $r$ is the distance from the GC. Usually such dependence is illustrated by choosing a
few different parametrizations of the profile with varying
cuspiness ({\it e.g.,} the peaked Navarro-Frenk-White profile \cite{NFW} as opposed to the cored isothermal \cite{isothermal} or Burkert profiles \cite{burkert}).   
$N$-body simulations and observations
of our own galaxy suggest a reasonable range of shapes for $\rho(r)$,
and it would be interesting to explore the dependence of the
constraints on the characteristics of $\rho(r)$  within such a
range.  One purpose of the present work is to fill this gap by
quantifying the strength of the IC constraint for a wide range of DM
density profiles.   We limit ourselves to the Einasto parametrization
\cite{Graham:2005xx,Navarro2008}, which we find to be useful
for this purpose since it  allows for continuous variation in the cuspiness and it has been used to fit the results of high-resolution
$N$-body simulations. 

\smallskip

The constraints also depend upon which leptons, and how many of them,
are assumed to be produced in the DM annihilations.  For example,
models that produce only $e^+e^-$ are more strongly constrained, for
a given DM mass $M$, than those that produce two $e^\pm$ pairs, since
the spectrum of the former is harder. This is a relevant distinction
since some of the  models that are most natural from the  particle physics
point of view  necessarily produce the $4e$ (or at least $2e$ plus
an invisible hidden sector gauge boson)
rather than the $2e$ final
state.  Likewise models that produce muons are more strongly
constrained than those that produce only electrons.   Moreover
models using the gauge kinetic mixing portal produce an admixture of 
$e^\pm$, $\mu^\pm$ and
$\pi^\pm$ rather than just a single kind of final state. 
Failing to recognize these distinctions, one could be misled into
thinking that the general class of models is excluded, even though
there may exist specific examples that satisfy all the constraints.  

Thus another goal of the present work is to explore in detail the
viability of a very specific set of annihilating DM models, which are
also theoretically motivated \cite{nima}-\cite{nonabelian}.  Namely we consider DM $\chi_i$
transforming under a nonabelian hidden sector gauge symmetry, taking
the simplest example of SU(2) and the lowest representations (doublet
or triplet) for the DM.  One of the dark gauge bosons $B_i$ is
assumed to mix with standard model hypercharge (and thus the photon)
through a dimension-5 gauge kinetic mixing operator.   This model
has a number of appealing features. The DM annihilates into dark
gauge bosons,  $\chi\chi\to BB$.   By assuming the gauge symmetry
breaks below the GeV scale, so the $B_i$'s have mass $\mu\lsim 1$
GeV, it is assured that only light leptons or $\pi^\pm$ are produced
by the gauge boson decays, and no antiprotrons that would be in
conflict with the PAMELA observations.  The annihilation cross
section is naturally enhanced by Sommerfeld effect 
\cite{sommerfeld,nima} in this model,
allowing it to have the correct relic density in the Early Universe,
even though the cross section in the galaxy must be larger by a
factor of order 100 than that needed for thermal freeze-out.

A distinctive feature of the model is that small mass splittings
$\delta M\sim$ 1 MeV between the DM states are generated at one
loop.   Such splittings can be important for making the model
consistent with direct detection constraints, by causing DM
scattering on nucleons to be inelastic and endothermic. This allows
the rate of DM interactions in detectors to be  sufficiently
suppressed despite having relatively large couplings to nucleons 
\cite{nima}.   The mass splittings can moreover lead to a significant
increase in the Sommerfeld enhancement factor \cite{slatyer}, which
effect we take into account here.  In addition, they can 
potentially help to explain the annual modulation observed by
DAMA/LIBRA through the inelastic DM mechanism (see for example
\cite{idm}), or the 511 keV excess observed by INTEGRAL \cite{spi}
using the excited DM mechanism \cite{FW}-\cite{newccf}, although we do
not explore these directions in the present work.

A further refinement we make compared to most previous
analyses is to  take into account the dependence of the Sommerfeld
enhancement on the position in the Galaxy, an effect that was first 
pointed out in ref.\ \cite{robertson}. The Sommerfeld enhancement
depends sensitively on the DM velocity, and both the DM velocity
dispersion and escape velocity depend upon $r$. We find that this
typically makes the enhancement larger near the galactic center where
the gamma rays that are most important for the IC constraint are
produced.  The constraint thus becomes harder to satisfy than in
models with a spatially constant boost factor, as is usually assumed.

Nevertheless, we find examples of models that are marginally able to
produce  leptons consistent with the PAMELA and Fermi observations,
and which are also barely consistent with the various gamma ray
constraints.  The preferred models have DM with $M\cong 800$ GeV and
gauge boson masses below 200 MeV, so that only electrons and no muons
or charged pions are produced in the annihilations.  However we find
that the Sommerfeld enhancement factor would always be {\it too
large} to achieve this concordance if the DM was present at the
expected relic density.  We are obliged to suppress the rate of
annihilations by introducing a factor $1/f$ in the density of the 
annihilating component.  This can be achieved by increasing the gauge
coupling appropriately so that the relic density is reduced.  The
minimum reduction $1/f=0.4$ occurs if the DM mass splitting is
negligible, and smaller values  $1/f \cong 0.14-0.2$ are needed if
$\delta M \sim 1$ MeV.  This comes about because of the $\delta
M$-dependence of the Sommerfeld enhancement factor for multistate
DM annihilations \cite{slatyer}.

In the present work we do not consider astrophysical boost factors
due to  increased annihilation rate in substructures of the main 
halo \cite{Kuhlen:2008aw}-\cite{Kamionkowski:2010mi}.  Inclusion of
these effects would presumably strengthen the gamma ray constraints
on the models,  but on the other hand a fully consistent treatment of
substructure effects should simultaneously consider the modification
to the  lepton signal from annihilations in subhalos
\cite{Brun:2009aj}.  This could conceivably have a compensating
effect that would weaken the constraints due to requiring less
$e^\pm$ production near the galactic center \cite{subhalo}.  Further
investigation of this issue is in progress.

We start by introducing the particle physics models in section
\ref{models}, and recalling the constraint upon the dark SU(2)
gauge coupling imposed by the relic density.   In section
\ref{boost_sect} we analyze how large an enhancement factor is needed for
the models to explain the anomalous lepton observations, the
theoretical computation of the Sommerfeld enhancement, and its phase
space average in the galaxy.  Section \ref{icconst} explains how
we implement the inverse Compton $\gamma$ ray constraint for general
DM Einasto profiles, while other relevant constraints are discussed in
section \ref{other}.  Our main results showing which parameter ranges
can be compatible with all the observations are presented in section
\ref{results}.  We discuss their implications in section
\ref{discussion}. Three appendices give details for the branching ratio of ground state 
DM annihilation into leptons (app.\ \ref{brfr}), the multistate
Sommerfeld enhancement (app.\ \ref{bfa}) and the calculation
of the DM escape velocity (app.\ \ref{escvel}) as a function of radial
position.

\section{Particle physics models}
\label{models}

\subsection{The models}
For the purpose of testing a concrete and specific model, 
we will analyze the simplest examples of nonabelian
hidden sector models that can explain the various cosmic ray anomalies;
namely a dark SU(2) gauge group under which the DM transforms as a
doublet or a triplet \cite{CCF, nonabelian}.  The Lagrangian for the DM and the gauge
bosons is
\beq
 {\cal L} = \sfrac12\bar\chi_i (i\slashed{D}_{ij}-M_\chi\delta_{ij})\chi_j 
-\sfrac{1}{4 g^2} B^a_{\mu\nu} B_a^{\mu\nu} - \sfrac{1}{\Lambda}\Delta_a
B_a^{\mu\nu}Y_{\mu\nu}
\label{lag}
\eeq
We assume that the dark Higgs triplet $\Delta_a$ gets a VEV in the
$a=2$ direction, so that $B_2$ mixes with the standard model 
hypercharge $Y_{\mu\nu}$ with strength 
\beq 
\epsilon =
\langle\Delta_2\rangle/\Lambda
\label{epsdef}
\eeq  
This causes $B_2$ to acquire a
coupling of strength $\epsilon q$ to any SM particle of charge $q$,
and therefore to mediate the decay $\chi_2\to\chi_1 e^+ e^-$
in the doublet model, or  $\chi_3\to\chi_1 e^+ e^-$ in the triplet
model.\footnote{The gauge couplings have the form
$g \bar\chi_i \sigma^a_{ij}\slashed{B}_a \chi_j$ or
 $\frac12 g \epsilon_{abc} \bar\chi_a \slashed{B}_b\chi_c$
respectively, in the doublet and triplet models.}
The value of $\epsilon$ is not strongly constrained; it can be
anywhere between $10^{-2}$ and $10^{-9}$ \cite{nonabelian} (except for a
small region around $10^{-6}$-$10^{-7}$ excluded by the E137
experiment if the dark gauge boson mass is less than 400 MeV
\cite{bjorken}).

These models can explain the PAMELA positron excess and
Fermi/LAT $e^\pm$ excess through the annihilation $\chi_1\chi_1\to
B_2 B_2$, followed by the decays $B_2\to \bar f f$, where $f$ is
any charged SM particle with mass less than $\mu_2/2$ (where $\mu_2$
denotes the mass of $B_2$).  Although higher mass fermions can be 
produced through off-shell gauge bosons, such processes have a much 
smaller cross section than those in which the $B_2$'s are on shell.
Using this fact, one can account for the lack of any 
antiproton excess in the PAMELA observations by assuming that $\mu_2$
is less than $2 m_p$.  

\subsection{Relic density}
\label{relic_sect}

Ref.\ \cite{nonabelian} computed the cross section for
$\chi\chi\to BB$ and determined that it leads to the right relic
density from  thermal freezeout in the doublet or triplet model, 
if the
gauge coupling $\alpha_g = g^2/4\pi$ has strength 
\beqa
\alpha_g =\left({M\over 1\hbox{\ TeV}}\right)
	\times\left\{\begin{array}{ll} 0.077, & \hbox{doublet}\\
0.031, & \hbox{triplet}\end{array}\right.
\label{relic_density}
\eeqa
corresponding to $g=0.98,\, 0.62$ respectively at $M=1$ TeV.  We will
adopt this relation between $g$ and $M$ in our analysis of  the
triplet model.  For the doublet model, the DM must be Dirac in order
to have a bare $M\bar\chi\chi$ mass term that respects the SU(2)
gauge symmetry.  It could therefore have a chemical potential that
fixes its relic density independently of its annihilation cross
section. This could provide motivation to consider $\alpha_g$ as an
additional free parameter for the doublet model, if one can find a 
mechanism for producing the asymmetry between $\chi$ and $\bar\chi$.

In later sections we will find  that it is difficult to satisfy all
constraints if the annihilating DM has the full allowed relic density,
because the Sommerfeld enhancement of the annihilation cross section
is too large.  A more general possibility is that the gauge coupling exceeds
(\ref{relic_density}) by a factor $\sqrt{f}$, which leads to an
enhancement in the cross section $\sigma\to f\sigma$, and a 
consequent suppression
in the density by $\rho \to \rho/f$.  Since the signals scale like $\rho^2\sigma$,
this leads to a reduction by $1/f$ in the rate of annihilations in the
galaxy.  Thus more generally we will consider gauge couplings given by
\beqa
\alpha_g = \sqrt{f}\left({M\over 1\hbox{\ TeV}}\right)
	\times\left\{\begin{array}{ll} 0.077, & \hbox{doublet}\\
0.031, & \hbox{triplet}\end{array}\right.
\label{relic_density_f}
\eeqa
where $f\ge 1$.  

If $f>2$, then less than half of the DM resides in the ground state,
and there must exist some additional DM sector to make  up the rest
of the mass density. If $1<f<2$, the needed reduction could happen
naturally within the model due to the fact that some fraction (less
than $1/2$) of the DM resides in one of the excited states, assuming
it is stable, and that it is not able to annihilate into leptons.
(Our triplet DM model provides such an example.)  A quantitative 
calculation of the relic density of the excited state is beyond the
scope of this paper, but will be carried out in ref.\ \cite{newccf}.

\subsection{Higgs sector}
\label{Higgs}

The model must also include a dark Higgs sector to break the
SU(2) gauge symmetry and give masses $\mu \lsim 1$ GeV to the
gauge bosons.  One-loop self-energy corrections to the DM masses
induce mass splittings of order $\delta M \sim \alpha_g\delta\mu$
if the gauge boson masses themselves are split by an amount of
order $\delta\mu$.  The splittings $\delta M\sim 1$ MeV are
potentially useful for explaining the INTEGRAL 511 keV gamma rays
by the excited dark matter mechanism, and they are needed to
satisfy constraints from direct detection unless the mixing 
parameter $\epsilon$ defined in (\ref{epsdef}) is small,
$\epsilon < 10^{-6}$ \cite{nima}.  Larger mixing $\epsilon\sim
10^{-3}$ are more interesting from the perspective of laboratory
tests by fixed-target experiments.

In the simplest and most predictive models, the DM states would have
no Yukawa couplings to Higgs bosons of the hidden sector.\footnote{In
a model with only triplet Higgses and Majorana DM, the only
potentially gauge invariant coupling
$\epsilon_{abc}\chi_a\chi_b\Delta_c$ in fact vanishes.}
 If one is
also interested in trying to address the 511 keV gamma ray excess,
the relation $\delta M = \frac12\alpha_g\delta\mu$ gives a
constraint on splittings between the gauge boson masses.   The DM
mass splittings should not be much more than of order 1 MeV; otherise
excited DM decays would produce $e^\pm$ of too high energy to be
compatible with the narrow observed 511 keV spectrum. For the triplet
DM model with $\alpha_g = 0.031$ for $M=1$ TeV, we thus expect that
$\delta\mu \lsim 180$ MeV if the largest DM mass splitting is $\delta
M = 3$ MeV.  Moreover we will show that models with $\mu\sim 200$ MeV
can more easily satisfy stringent $\gamma$ ray constraints.

\subsection{Explicit example}
\label{explicit}

Ref.\ \cite{nonabelian} showed that the minimal Higgs sector for
splitting triplet DM is one that gives a VEV to an additional
triplet Higgs field $\Delta_a'$ in a direction orthogonal to that
of the first triplet VEV, {\it e.g.,} $\Delta_1'$.  We do
not wish to commit ourselves to a particular Higgs potential in
the present analysis. Instead, we will assume that it is possible
to design one that gives the desired DM mass splittings of order
100 keV to a few MeV, and gauge boson masses of order a few 100
MeV.  In fact, this is quite reasonable even in the context of
the simplest model with two Higgs triplets.  Using the analysis
of ref.\ \cite{nonabelian}, the mass splittings $\delta M_{13} =
3.1$ MeV, $\delta M_{23} = 100$ keV and gauge coupling $g=0.62$
imply that $\Delta_1' = 320$ MeV,  $\Delta_2 = 80$ MeV, and the
gauge boson masses are $\mu_1 \cong 50$ MeV, $\mu_2\cong \mu_3
\cong 200$ MeV using this simplest Higgs sector.

\section{Sommerfeld enhancement for PAMELA/Fermi lepton signals}
\label{boost_sect}

\begin{figure}[t]
\bigskip \centerline{
\epsfxsize=0.5\textwidth\epsfbox{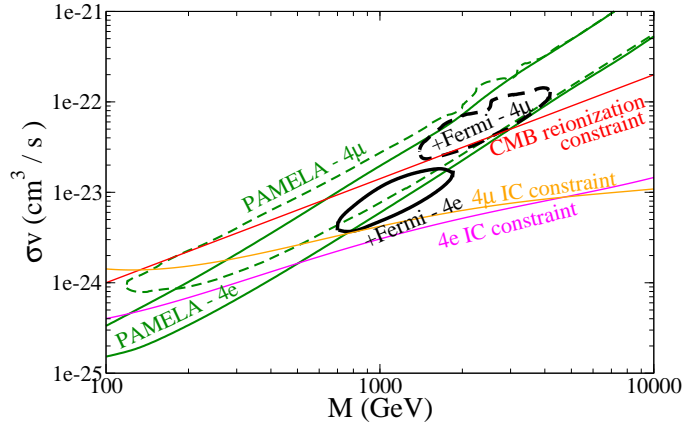}}
\caption{Heavy lines: allowed regions for explaining the PAMELA and Fermi lepton
excesses, assuming $4e$ or $4\mu$ final states, from ref.\
\cite{papucci}.  The fits assume an Einasto DM profile with
$\rho_\odot = 0.3$ GeV/cm$^3$, $\alpha=0.17$ and $r_s = 20$ kpc.
Narrow lines show constraints from CMB
\cite{CIP} and inverse
Compton gamma rays \cite{papucci}.}
\label{allowed}
\end{figure}

In this section we explain how we calculate the effective cross
section for annihilating DM to produce high-energy $e^\pm$ (and
possibly heavier charged particles) in the galaxy, relevant for
explaining the PAMELA and Fermi $e^\pm$ excesses.   A particularly
simple case to focus upon is that where the mass $\mu$ of the $B_2$
gauge boson  is below the threshold for producing a $\mu^+$-$\mu^-$
pair.  In that case one need only consider final states consisting of
$e^\pm$.  Moreover the cross section to fit the PAMELA/Fermi lepton
data is smaller in this case, and thus easier to achieve from the 
model-building perspective.  Another reason for preferring the $4e$
channel is that it is not ruled out by CMB reionization constraints,
whereas the $4\mu$ channel is (depending upon the assumed value of
the local DM density $\rho_\odot$, as we will discuss); this is
apparent in figure \ref{allowed}.

\subsection{Cross section ratio for $\chi_1\chi_1\to e^\pm$}

Although the models we consider could have an excited DM state that
is stable or metastable, the ground state normally should have a
larger density $n_1$ due to downscattering  processes like
$\chi_2\chi_2\to\chi_1\chi_1$ in the early universe.  To simplify the
analysis, we assume that it is a  good approximation to ignore
annihilations involving the subdominant component for the production
of leptons via $\chi\chi\to 4e$.  For the triplet DM model, there are
in fact no such annihilations in the form $\chi_2\chi_2\to 4e$
because $\chi_2$ does not couple to $B_2$, the gauge boson that we
have assumed to be the only one mixing with the SM.  The only
annihilations we miss in this case are of the form $\chi_1\chi_2\to
B_1 e^+ e^-$, where the $B_1$ appears as missing energy.

It is potentially important to realize that only the ground state 
(and
possibly some small addition of excited state particles)
contribute to the annihilation in the galaxy, because the cross
section for $\chi_1\chi_1$ to annihilate is different from that for
$\chi_i\chi_j$ in the early universe.  They are related by the ratio
$b_{11}$,
\beq
	\sigma(\chi_1\chi_1\to e^\pm) = b_{11} 
	\sigma(\chi_i\chi_j\to {\rm any})
\label{sigma_part}
\eeq
which in the doublet and triplet models  is given by
\beq
 b_{11} = \left\{\begin{array}{ll} \frac59, & \hbox{doublet}\\
 & \\
\frac{8}{7}, & \hbox{triplet}\end{array}\right.
\eeq
as we explain in appendix \ref{brfr}.\footnote{we thank Tracy Slatyer for pointing out an error in our
original computation of this quantity}

\begin{figure*}[t]
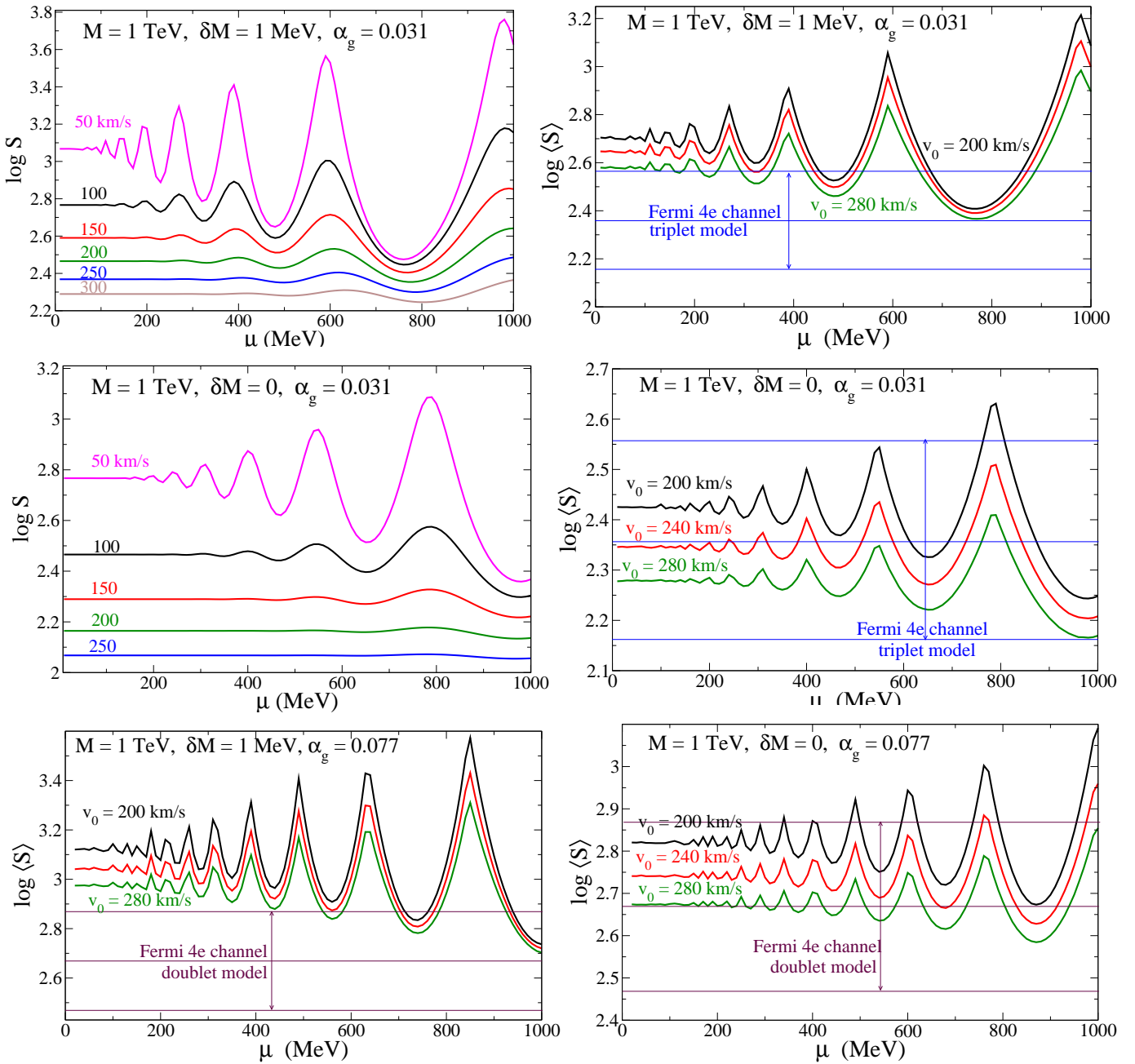

\smallskip \centerline{
\epsfxsize=0.48\textwidth\epsfbox{mu-v-dep.eps}
\epsfxsize=0.52\textwidth\epsfbox{averaged2.eps}}
\smallskip \leftline{
\epsfxsize=0.49\textwidth\epsfbox{mu-v-dep-dM0.eps}
\epsfxsize=0.51\textwidth\epsfbox{averaged-dM02.eps}}
\smallskip \leftline{
\epsfxsize=0.5\textwidth\epsfbox{avg-doublet-dM.eps}
\epsfxsize=0.5\textwidth\epsfbox{averaged-doub-dM02.eps}
}
\caption{Upper left: Sommerfeld enhancement factor (\ref{slat}) as a function of gauge
boson mass $\mu$ for DM mass $M=1$ TeV, mass splitting 
$\delta M = 1$ MeV,  and gauge coupling $\alpha_g = 0.031$.  Different
curves are labeled by DM velocity in center of mass frame, in 
km/s.  Upper right: the same, but averaged over the DM velocity distribution
functions as in (\ref{vavg}), for three different circular
velocities $v_0 = 200$, 240 and 280 km/s.  The allowed regions for
Fermi with $4e$ annihilation channel for the 
doublet and triplet DM models, assuming solar density
$\rho_\odot = 0.3$ GeV/cm$^3$ and DM fraction $f=1$ (see eq.\ 
(\ref{relic_density_f})) 
are shown by horizontal lines.
Central panels: same as upper but with vanishing mass splitting
$\delta M = 0$.
Lower panels: velocity-averaged Sommerfeld enhancement factors 
versus $\mu$ for
doublet model, with $\alpha_g = 0.077$ and $\delta M= 1$ MeV
(left), $\delta M=0$ (right).
}
\label{boost}
\end{figure*}

\subsection{Sommerfeld enhancement for multistate DM}

The boost factor needed for explaining the PAMELA/Fermi $e^\pm$
observations can be inferred using figure \ref{allowed}, which
shows the $3\sigma$-allowed regions for $M$ and $\sigma v$,
assuming $4e$ and $4\mu$ final states, and also assuming the solar
neighborhood DM density to be $\rho_\odot = 0.3$ GeV/cm$^3$.
The central value of the
$4e$ allowed region has $\sigma v = 10^{-23.1\pm 0.2}$ cm$^3$/s at 
$M\cong 1.1$ TeV, while the $\sigma v$  needed for 
the right relic abundance is $3\times 10^{-26}$ cm$^3$/s.  This
leads to a required boost factor of 
$B = {10^{-23.1\pm 0.2}/( b_{11} \cdot 3\times 10^{-26})}$.
Allowing for $\sigma v\to f\sigma v$ and $\rho\to \rho/f$ as 
described in section \ref{relic_sect}, the required boost factor
scales as $B\to f B$.  
For the doublet and triplet models, using the gauge couplings
(\ref{relic_density_f}), this gives
\beq
	\log B_{4e} = \log f + \left\{\begin{array}{ll} 
	2.67 \pm 0.2,& \hbox{doublet}\\
	2.36 \pm 0.2,& \hbox{triplet}\end{array}\right.
\eeq
for the boost factor needed in the $4e$ annihilation channel.

To get a feeling for the ability of the models to achieve large
enough boosts, we start by  
computing the Sommerfeld enhancement factor for a fixed DM mass
of $M=1$ TeV near the central region allowed by Fermi with 4$e$
final states.  
For this purpose one should take into account that
there is more than one DM state, so the usual Sommerfeld factor
for single-state DM is not appropriate.  The general case has
not been solved explicitly, but that of two states with therefore
a single mass splitting was recently
analyzed in \cite{slatyer}, where it was found that the enhancement
factor can be approximated by eq.\ (\ref{slat}) given in appendix
\ref{bfa}.

The result (\ref{slat}) applies directly to the doublet model, where
there are only two states, but only approximately  in the case of the
triplet model, since the latter has three states and two mass
splittings.  Since the enhancement is an increasing function of
$\delta M$, one might reasonably expect the  multistate effect in the
triplet model to be captured by choosing the largest of the two mass
splittings.  We make this assumption, which in a more ambitious study
should be checked.  It is however a technically challenging problem. 

\subsection{Velocity-averaged enhancement at $r=r_\odot$}

We must compute the boost factor in the neighborhood of the Sun 
in order to predict the rate of high-energy lepton production from
DM annihilations.
The Sommerfeld enhancement depends upon the relative velocity of the DM
particles, which is usually characterized by a Maxwellian
distribution 
\beq
	f(v) = N e^{-v^2/v_0^2}\,\theta(v-v_{\rm esc})
\eeq
with a cutoff for $v$ above some escape velocity $v_{\rm esc}$.
The value $v_0$ is commonly taken to be 220 km/s at the solar radius
$r_\odot =8.3$ kpc, although 
higher values $\sim 250$ km/s have been advocated more recently 
\cite{highv}.  The range $v_0 \in 200-280$ km/s is
suggested by the compilation of different measurements in ref.\ 
\cite{vrange}.
The escape velocity is directly correlated with $v_0$, as detailed
in appendix \ref{escvel}.  There we motivate our choice for the
$r$-dependent relation
\beq
	v_{\rm esc}^2(r) = 2 v_0^2(r)\left[2.39 + 
\ln({\rm 10\ kpc}/r)\right]
\label{vesc}
\eeq
in the region $r < 10$ kpc. (We elaborate on the $r$ dependence of
$v_0$ in section \ref{boostr}.) At the solar radius, which we take to 
be $r_\odot = 8.33$ kpc \cite{solar_radius}, this gives
$v_{\rm esc} = 2.53\, v_0$, corresponding to 
escape velocities in the range $440-610$ km/s.  This is in reasonable
agreement with constraints from direct measurements of high-velocity
stars \cite{rave}.
We need to average $S(|\vec v_1-\vec v_2|)$ over the
phase space,
\beq
	\langle S \rangle = \int d^3 v_1 d^3 v_2 f(v_1) f(v_2)
	S(|\vec v_1-\vec v_2|)
\label{vavg}
\eeq

\subsection{Predicted versus desired enhancement factor values}

In figure \ref{boost} the dependence of the Sommerfeld enhancement
$S$ and its velocity average $\langle S\rangle$ on the gauge boson
mass $\mu$ is shown, for the example of $M=1$ TeV and for the triplet
model with $\alpha_g=0.031$, as well as the doublet with 
$\alpha_g=0.077$.  These are the gauge couplings needed for the right
relic density in the respective cases.  The extra enhancement due to
the mass splitting $\delta M = 1$ MeV is quite significant; in fact
it tends to give rise to boost factors that are too large compared to
the values needed to explain the high-energy lepton excesses.   
The effect of $\delta M$ actually 
saturates around $\delta M = 600$ keV for the triplet model and
$\delta M = 100$ keV for the doublet model; the enhancement remains
roughly constant for larger
values (however the validity of the approximations leading to
(\ref{slat}) breaks down if $\delta M \gsim \alpha_g^2 M$).

We conclude that, from the point of view getting the right boost
factor, there is no need for the mass splitting, and in fact too large
a mass splitting is disfavored.  However this can be compensated by
invoking the $f$-factor of section \ref{relic_sect}, since taking
$f>1$ increases the value of the needed boost factor.  

\section{Inverse Compton gamma ray constraint}
\label{icconst}

To undertake a thorough exploration of the parameter space, we must
take into account constraints that could rule out the
models.   The intepretation of the PAMELA and Fermi excesses as being
due to DM annihilation has come under pressure from a number of
complementary constraints. Many of these arise from radiation that would be
produced by the leptons after annihilation, which should be directly
detectable as gamma rays or radio emission in the galaxy, or
indirectly in distortions of the cosmic microwave background by
contributing to the reionization of the early universe.  The most
stringent constraint on our model is from inverse Compton scattering
in the galaxy.  We therefore consider it separately in this section,
and discuss other constraints in section \ref{other}.

\begin{figure*}[t]
\smallskip \centerline{\epsfxsize=0.75\textwidth\epsfbox{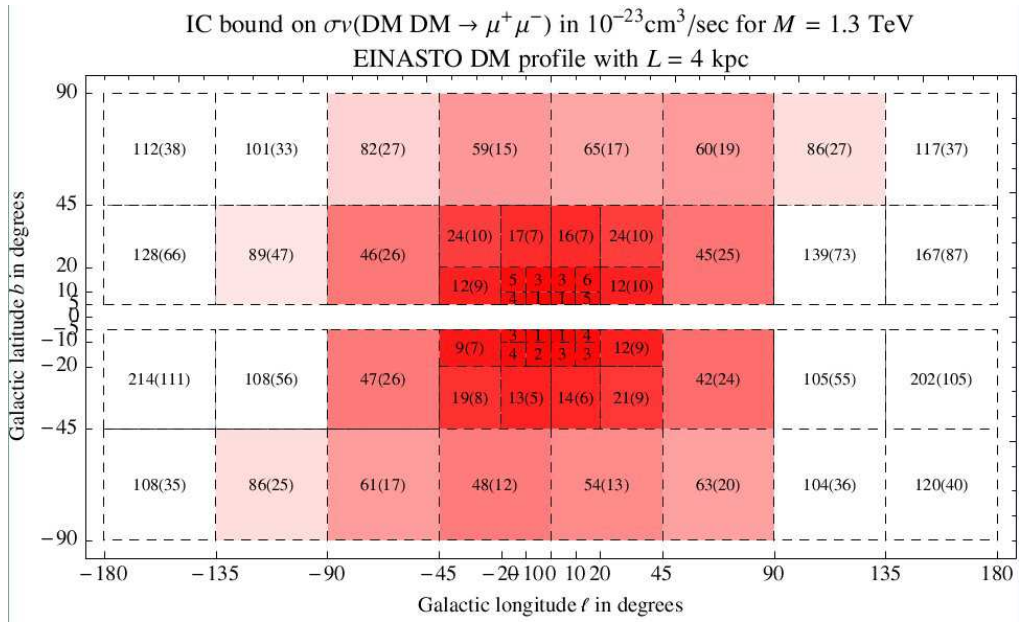}}
\caption{Regions near galactic center used by ref.\ \cite{papucci}
for computing the
Fermi IC bound. Numbers in each region refer to constraints arising for a particular
annihilation channel (into $\mu^+\mu_-$), but size and shape of
regions are the same for the channel we consider, $2e^+2e^-$.
 Courtesy of A.\ Strumia.}
\label{skymap}
\end{figure*}

\subsection{Characterization of the bound}

If DM annihilates into charged particles, the latter will undergo
inverse Compton scattering on the galactic radiation field (for
example, starlight) to produce high energy gamma rays that could be
detected by the Fermi LAT.  Several authors have considered the
constraints arising from the nondetection of such a signal
\cite{CP,cuoco,Regis:2009md,Cholis:2009gv,papucci,CPS,zaharijas,hutsi}.  Most of
these assume a two-lepton final state rather than four leptons, as is
the case in our models.  The constraints on the former are stronger
because their spectrum is harder than in the four-body case. Ref.\
\cite{papucci} has considered four-lepton final states for
several DM density profiles, so we adopt their results for the present
analysis.

\begin{figure}[t]
\smallskip \centerline{
\epsfxsize=0.5\textwidth\epsfbox{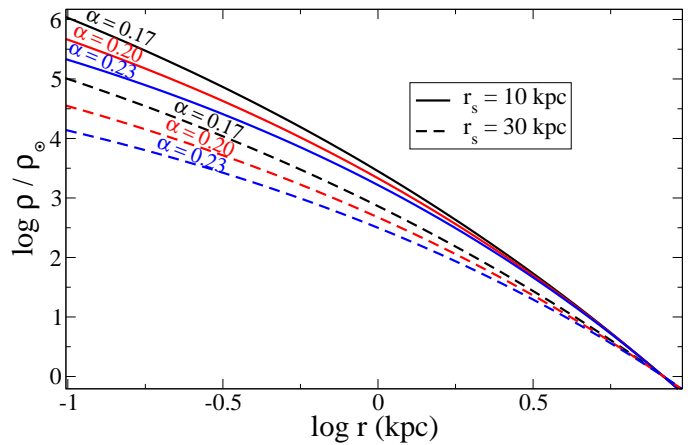}}
\caption{Einasto density profiles normalized to $\rho_\odot$ for $\alpha = 0.17,0.20,0.23$
and $r_s = 10$ kpc (solid curves) or $r_s = 30$ kpc (dashed curves)}
\label{logrho}
\end{figure} 

\begin{figure*}[t]
\smallskip \centerline{\epsfxsize=0.8\textwidth\epsfbox{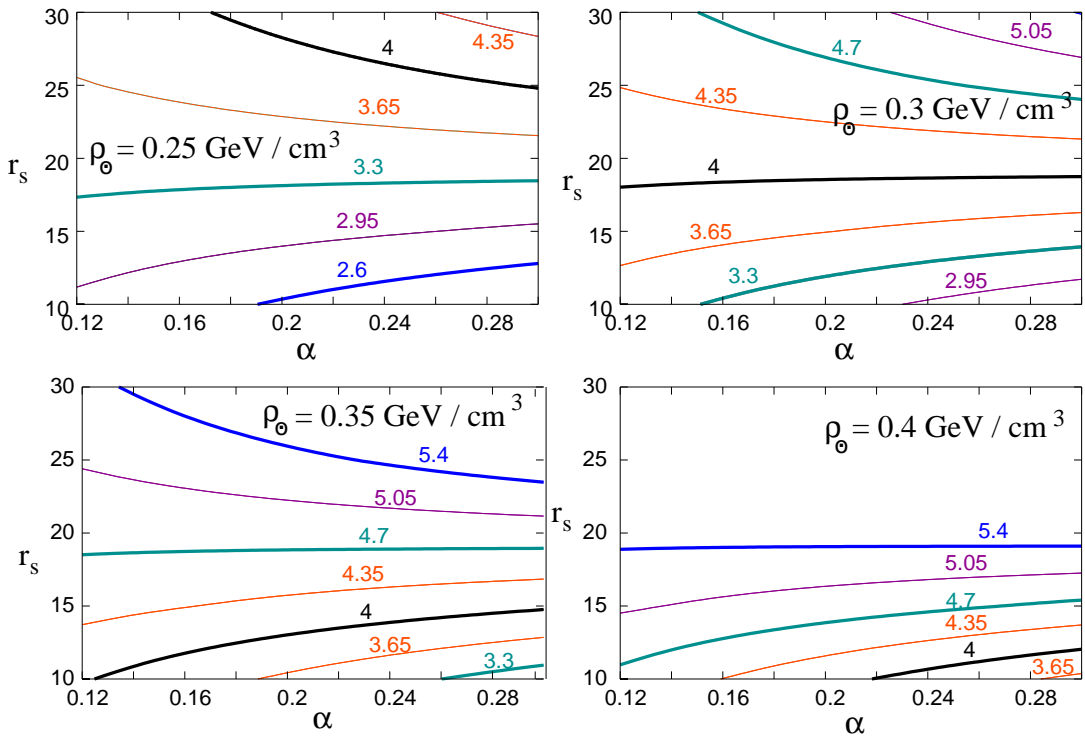}}
\caption{Contours of $M_{60}/(10^{11} M_\odot)$ 
($M_{60}$ being the mass contained within 60 kpc) 
in the plane of $\alpha$ and $r_s$ (in kpc units), for several values
of $\rho_\odot$.  The heavy
line labeled ``4'' is the central value,
and other heavy lines (at $4\pm 0.7$ and $4\pm 1.4$)  denote the neighboring $1\sigma$ and $2\sigma$
confidence intervals determined by ref.\ \cite{xue}.}
\label{xuecont}
\end{figure*}

The resulting fits of \cite{papucci} for the PAMELA/Fermi
excesses and the IC constraints for the $4e$ and $4\mu$ channels are
shown in figure \ref{allowed}, for a particular choice of 
Einasto profile,
\beq \rho(r) = \rho_\odot \exp\left[
-{2\over \alpha}\left( \left({r\over r_s}\right)^\alpha - 
\left({r_\odot\over r_s}\right)^\alpha\right)\right] 
\label{fiducialEinasto}
\eeq 
 with $\rho_\odot = 0.3$ GeV/cm$^3$, $\alpha = 0.17$,
$r_s = 20$ kpc, and $r_\odot = 8.5$ kpc.    The maximum allowed annihilation cross section
for these two cases can be approximated as\footnote{tables on which these
fits are based provided by A.\ Strumia} 
\beqa
4e:  \log \sigma_{4e} v &=& -15.14 - 14.66\, x + 8.00 \,x^2 - 1.78\,
x^3 \nonumber\\ &+& 0.144\, x^4  \nonumber\\
4\mu:   \log \sigma_{4\mu} v &=&
-8.245 - 21.37\, x + 10.43\, x^2 - 2.136 \,x^3\nonumber\\ 
&+& 0.161\, x^4
\label{bounds}
\eeqa
where $\sigma v$ is in units of cm$^3$/s and $x = \log M$ with $M$
in GeV.  The constraints are found by comparing the observed 
gamma ray spectrum from Fermi data with the predicted IC contribution 
from the DM-produced leptons, and demanding that the latter do not
exceed the former in any energy bins by more than 3$\sigma$.  
Ref.\ \cite{papucci} carried out this procedure
for different regions of the sky, similar to those
depicted in fig.\ \ref{skymap}.  The  regions closest to the galactic
center are the ones that give the strongest constraint
on the cross section, resulting in the bounds (\ref{bounds}).
It is worthwhile to notice that the inner $5^\circ$ of galactic
latitude have not been used in deriving the constraints, due to the
difficulties of interpreting foregrounds associated with the disk of the 
galaxy.  
We have adopted ref.\ \cite{papucci}'s constraints relative to
the `MED' propagation model for $e^\pm$, corresponding to a 
diffusion zone thickness
of $L=4$ kpc; the results from other choices
(except for the extreme value $L=1$ kpc)
are not significantly different in the range of DM masses in which we
are interested.

For realistic models of the type we consider, if the gauge boson
is heavy enough to decay into muons, then electrons will also
necessarily be produced with some nonnegligible probability, depending
only on the relative phase spaces for decays into the two different
final states.  Moreover charged pions will also be produced if
the gauge boson mass $\mu$ is greater than $2 m_\pi$.   In such cases
the actual bound will be somewhere in between $\sigma_{4e}$ and
$\sigma_{4\mu}$ given in (\ref{bounds}).  We will avoid the
uncertainty of how to correctly interpolate between the two bounds
by restricting our attention to the case $\mu < 2 m_\mu$, so that
only the $4e$ bound applies.

\subsection{Dependence of IC bound on DM profile}

Figure \ref{allowed} shows that for the assumed fiducial density
profile (\ref{fiducialEinasto}), the constraints rule out the Fermi-allowed regions.  This
can be circumvented however by considering DM density profiles that
are less strongly peaked near the galactic center, where the most
stringent constraints come from.  With the Einasto profile, both
$\alpha$ and $r_s$ have an effect on  the cuspiness of $\rho(r)$ near
the galactic center, which is  illustrated in figure \ref{logrho};
larger values of $\alpha$ or $r_s$ weaken the bound.  One of our
goals is to quantify this statement to determine the range of
$\alpha$ and $r_s$ that gives a consistent description of the
PAMELA/Fermi anomalies in terms of DM annihilation.

\subsubsection{Theoretical and observational constraints on the DM
profile}

Let us first consider what are the reasonable ranges of variation for
the halo profile parameters $\alpha$ and $r_s$  based on theoretical and observational
considerations.  $N$-body simulations like Aquarius
find $0.115 < \alpha < 0.179$ \cite{Navarro2008}; in a different fit of
the same galaxies over a slightly smaller radial region, ref.\
\cite{tissera} finds $\alpha$ as large as $0.19$.  These values are
for pure DM simulations without baryons; the inclusion of baryons 
tends to reduce $\alpha$ dramatically to the range $0.07-0.015$ due to the
concentrating effect of the baryons on the inner halo
\cite{tissera,abadi,pedrosa}.   In contrast,  the pure DM simulations find a larger
range of $r_s$, from 15 kpc to 29 kpc (where we have assumed $h=0.7$
for the Hubble parameter, since the results are quoted in units of
kpc$/h$).  This range also gets decreased when baryons are included,
to $5-15$ kpc.  There thus appears to be a  conflict between the DM
simulations with baryons (BDM) and the annihilating DM interpretation
of PAMELA/Fermi events, since the adoption of $\alpha=0.15$ and
$r_s=15$ kpc will only make the IC constraint stronger, compared to
values that already rule out the model.  We also note that one of the
highest resolution simulations, GHALO \cite{ghalo}, obtains $\alpha=0.155$
even without including baryons.

For the allowed range of the solar neighborhood density $\rho_\odot$,
observations provide tighter constraints than do simulations.  The
value $\rho_{\odot,0} \equiv 0.3$ GeV/cm$^3$ has for a long time been
considered standard \cite{pdg}. More recently a higher central value 
$0.385\pm  0.027$  GeV/cm$^3$ has been advocated in
\cite{CatenaUllio} and  $0.43\pm 0.15$ GeV/cm$^3$ has been determined
using a method that does not rely upon the detailed form of $\rho(r)$
in \cite{salucci}.


Observational limits on the ranges of the other Einasto parameters
tend to be weaker than the ranges suggested by simulations, but
this depends upon the assumed value of  $\rho_\odot$.
One constraint comes from the mass within a given radius inferred from
rotation curves of the Milky Way galaxy.  Ref.\ \cite{sakamoto}
inferred a total mass within 50 kpc of the galactic center,
$M_{50} = 5.3\times 10^{11}\, M_\odot$.  More recently, using data
from the Sloan Digital Sky Survey, ref.\ \cite{xue} obtained the
mass within 60 kpc as $M_{60}= (4\pm 0.7)\times 10^{11}\, M_\odot$.
In fig.\ \ref{xuecont} we plot contours of $M_{60}/(10^{11} M_\odot)$
in the $\alpha$-$r_s$ plane, for several values of $\rho_\odot$.  At
$\rho_\odot=0.4$ GeV/cm$^3$, values of $r_s$ exceeding 20 kpc are
disfavored by this measurement, while at $\rho_\odot=0.3$ GeV/cm$^3$
it imposes essentially no constraint.

\begin{figure}[t]
\smallskip \centerline{\epsfxsize=0.4\textwidth\epsfbox{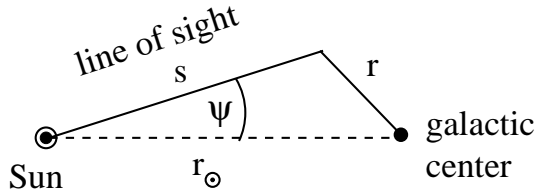}}
\caption{Representation of the quantites $r$, $s$, $\psi$ used to compute
the $\bar J$ factor (\ref{Jeq}). The coordinate $s$ runs along the line of sight in the 
direction
specified by the angle $\psi$ and is related to $r$ by
$r(s,\psi)=\sqrt{r_\odot^2+s^2-2r_\odot\ s\, \cos\psi}$. }  
\label{geometry}
\end{figure}

\subsubsection{$\bar J$ factors for different density profiles}

A principal goal of this work is to investigate the sensitivity of the IC bounds
in (\ref{bounds}) to different choices of the Einasto profile
parameters.  The dependence upon $\rho_\odot$ is easy to quantify,
since the rate of DM annihilations is proportional to $\rho^2\sigma$.
This implies that for a general value of $\rho_\odot$, and the 
fraction $1/f$ of annihilating DM defined through
(\ref{relic_density_f}),  the upper bound 
on the annihilation cross section scales as
\beq
	\sigma_{4e} \to \sigma_{4e} \left({\rho_{\odot,0}\over
	\rho_\odot}\right)^2\, f^2
\label{s4es}
\eeq
relative to that at the reference solar density $\rho_{\odot,0} = 0.3$
GeV/cm$^3$ and $f=1$.
This means that the constraints become more severe by a factor of 
$(0.43/0.3)^2 = 2$ if we adopt the new central value
$0.43$ GeV/cm$^3$ of ref.\ \cite{salucci} instead of the
value $0.3$ GeV/cm$^3$ assumed in ref.\ \cite{papucci}.
However one must remember that the cross section required to fit
the PAMELA and Fermi leptons goes down by the same factor, so in
fact the value of $\rho_\odot$ has little direct effect on the ability of
a model to satisfy the IC constraint.\footnote{Indirectly it has an
effect through the constraint from $M_{60}$, which makes large
values of $r_s$ hard to achieve when $\rho_\odot$ is large.}

\begin{figure}[t]
\smallskip \centerline{\epsfxsize=0.5\textwidth\epsfbox{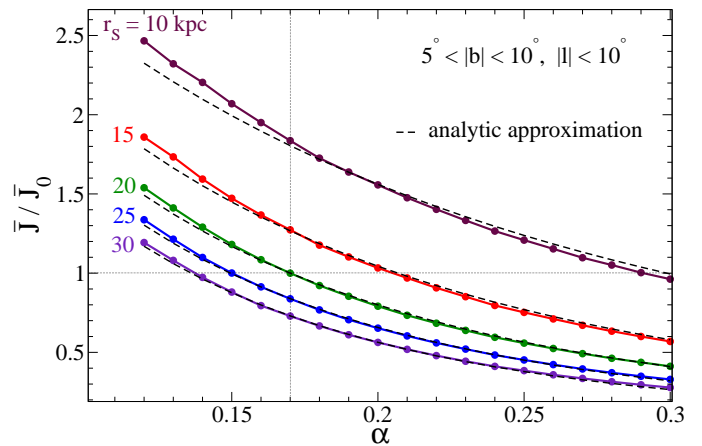}}
\caption{$\bar J$ factor eq.\ (\ref{Jeq}) over the fiducial value used
by ref.\ \cite{papucci} as a function of $\alpha$, for several values
of $r_s$.  Dashed curves are the analytic approximation
(\ref{approx}).}  
\label{Jvalpha}
\end{figure}

The effect of changing the shape of the profile, on the other hand, 
is harder to quantify.  The correct procedure would require solving
the diffusion equation for electrons in the galaxy again for each
choice of $\alpha$ and $r_s$, and then recomputing the spectrum of
IC scattering.  Here we content ourselves with a
simpler method of estimation, in which we ignore the changes to
the electron diffusion induced by changes in the shape of the
profile.  In this approximation, the rate of IC scattering due to
leptons originating from the DM annihilations changes only by a
geometrical factor, the averaged $\bar J$ factor, defined as
\beq
	\bar J = {1\over \Delta\Omega}\int d\Omega \int_{\rm l.o.s.}
{ds\over r_\odot}
	\left({\rho(r(s,\psi))\over \rho_\odot}\right)^2
\label{Jeq}
\eeq
where the integral along the line of sight is then averaged in 
$\psi$ (defined in fig.\ \ref{geometry}) over the solid angular regions $\Delta \Omega$
that are used to compute the gamma ray constraints, as shown in 
fig.\ \ref{skymap}.  As noted above, the relevant regions for obtaining
the strongest bound are those with latitude $5^\circ < |b| < 10^\circ$
and longitude $|l| < 10^\circ$. 
 The
direction $\psi$ corresponds to latitude $b$ and longitude $\ell$ as
$\cos\psi = \cos b \cos \ell$.

The resulting $\bar J$ factors, divided by the fiducial one $\bar J_0$ corresponding
to the parameters chosen by ref.\ \cite{papucci}, are shown as a
function of $\alpha$ for several values of $r_s$ in fig.\
\ref{Jvalpha}.  The IC upper bound on the annihilation cross section
is expected to scale like $\sigma_{4e} \to \sigma_{4e}/(\bar J/\bar J_0)$, 
so that smaller values of $J/J_0$ correspond to a weaker bound.
One observes that it is not easy to weaken the bound while remaining
within the expectations of $N$-body simulations.  To achieve a factor
of 2 reduction would require $\alpha\simeq 0.22$ and $r_s=30$, for example,
which appear to be extreme values. 

One can get analytic insight into the dependence of $\bar J$ on $\alpha$
and $r_s$ by hypothesizing that $\bar J$ scales like 
\beq
	\frac{\bar J}{\bar J_0} \cong \left({\rho_{0}(r_{\rm IC})\over
	\rho(r_{\rm IC})}\right)^2
\label{approx}
\eeq
where $\rho_0$ is the Einasto profile using the fiducial $\alpha=0.17$
and $r_s=20$ kpc values, and $r_{\rm IC}$ is some characteristic 
radius that should be of order
$r_\odot$ times the angular displacement ({\it i.e.,\ }5$^\circ$) from the galactic center of 
the relevant solid angular region.  By tuning $r_{\rm IC}$ to the 
value 1.75 kpc, we are able to get good agreement with the numerical
results, with less than 4\% error in the region $\alpha > 0.15$
(and less than 6\% elsewhere).  The approximation is shown as the
dashed curves in figure \ref{Jvalpha}.

\subsubsection{Potential for strenghtening of IC constraint}

\begin{figure}[t]
\smallskip \centerline{\epsfxsize=0.5\textwidth\epsfbox{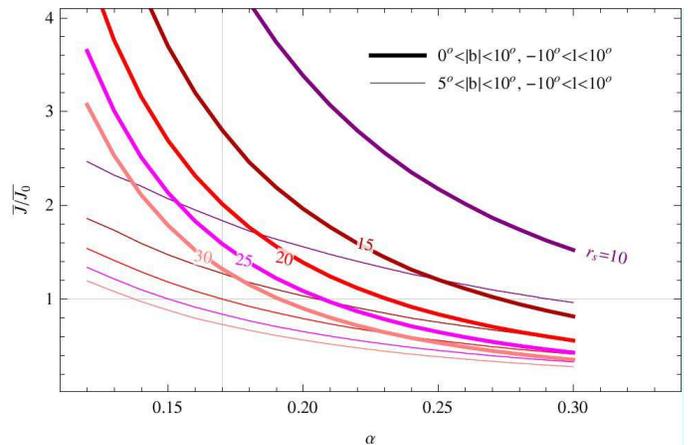}}
\caption{Solid curves: $J$ factors as in fig.\ \ref{Jvalpha}, but
averaged over the angular region including the inner latitudes
of the galaxy.  Dashed curves are for the original region considered
in figure \ref{Jvalpha}, for comparison.}
\label{jva5}
\end{figure}

As we noted above, the current constraints are based on a region that
excludes the central latitudes $|b| < 5^\circ$, due to the complexity
of the disk of the galaxy.
On the other hand we know that the strongest bounds arise from the
innermost regions.  It is therefore interesting to try to project how
much stronger the constraints might become if one had made use
of the Fermi/LAT $\gamma$ ray data from the inner latitudes.  It is
straightforward to recompute the $J$ factors over regions that include
the inner $\pm 5^\circ$ of latitude to try to project how much
stronger the bounds might become.  We show the result in fig.\ 
\ref{jva5}.  According to this extrapolation, the
bound could get stronger by a factor of 2 at the reference values
of $\alpha=0.17$ and $r_s=20$ kpc.  However this is merely suggestive,
and one would have to analyze the actual Fermi $\gamma$-ray data in
the inner region to draw firm conclusions.

\section{Other constraints}
\label{other}

As mentioned above, a number of complementary constraints have been
studied pertaining to models of annihilating DM. We present here the most
significant of them, and discuss how they apply to the cases in which
we are interested and their impact on the parameter space.

\subsection{GC gamma ray constraints}

Fluxes of gamma rays are directly produced by the DM annihilation
process itself (`prompt' gamma rays), mainly from the bremsstrahlung
of charged particles and the fragmentation of hadrons, {\it e.g.,}
$\pi^0$, produced in the annihilations. These fluxes extend to very
high energy (up to the mass of the annihilating DM particle) and are
therefore better constrained by comparing the predicted flux with the
observations of high energy gamma ray telescopes such as the 
H.E.S.S.\ or VERITAS (Very Energetic Radiation Imaging Telescope
Array System)  observatories. Refs.\ \cite{prompt1} and
\cite{meade,papucci} have indeed carried out the analysis of these
constraints on the parameter space of models that aim to explain the
lepton anomalies. H.E.S.S.\ observations of the region of the
Galactic Center \cite{HESSGC} (defined as a disk of radius
$0.1^\circ$ centered at the GC), of the Galactic Ridge \cite{HESSGR}
(defined as a region of $0.8^\circ \times 0.3^\circ$ in longitude and
latitude centered at the GC) and of satellite dwarf spheroidal
galaxies such as Sagittarius, Draco, Ursa Minor or Willman 1
\cite{dSph} impose constraints that can be significant (see
\cite{prompt1,meade,papucci}). 

However, all of these constraints are very sensitive to the details
of the chosen DM profile. In particular, for the Milky Way ones, it is the inner part of
the galactic DM halo which counts, and numerical simulations do not provide
direct determinations of the profile at $r \sim 100$ pc. Moreover,
assuming that an extrapolation is possible, only for the steepest of
the profiles that we consider (small $\alpha$ values) do these bounds
become competitive with those from IC, and such values are anyway
disfavored by the more robust IC constraints. 
We therefore need not consider these bounds any further.

\subsection{GC radio constraint}

The $e^\pm$ produced by DM annihilations within the galactic magnetic
field emit synchrotron radiation, which falls in a range of
frequencies roughly spanning the radio to the IR. The Galactic Center
is the best region to search for this effect, both because of the large
local value of the  DM density and of the magnetic fields. We do not enter
here into the details of the needed astrophysical and particle
physics ingredients  but refer to \cite{prompt1} for the complete
discussion. There, in brief, the signal is computed neglecting
advection and diffusion but scanning different (extremal) assumptions
for the galactic magnetic field. Ref.\ \cite{meade} has in particular
considered the case of the 4-lepton annihilation final state that we
interested in.

Comparing the predicted flux with observations produces constraints
on the DM annihilation cross section. Since the observed GC microwave
spectrum is harder than what DM annihilations can produce, the
dominant bound is obtained considering the observation available at
the lowest observed frequency, $\nu=0.408\,{\rm GHz}$, performed
by~\cite{Davies} in a region with full width half maximum of $4''$,
corresponding to about 0.1 pc. The resulting bounds can be quite
stringent (see \cite{meade}) and the constraint extends to low DM
masses where the $\gamma$-ray bounds from H.E.S.S., discussed above,
are not effective. The variation of the magnetic field negligibly
affects the bound, because the radio emission is predominantly
produced by outer regions. A subdominant bound comes from the VLT
observation~\cite{VLT} at the larger infrared/visible frequency from
a region with even smaller angular size $0.04''$ {\it i.e.,}
$r<0.0016\,{\rm pc}$. This bound somewhat depends on the magnetic
field profile, and it becomes numerically significant only for spiked
DM density profiles~\cite{UllioRegis}.

However the same discussion as above applies to these bounds: as for the GC gamma ray constraints, these bounds are very sensitive to the details of the chosen DM profile  (they originate from the even smaller regions of $r \sim 1$ pc around the GC) and do not apply to non-steep profiles. We therefore need not consider these bounds any further.

\subsection{Reionization constraint from CMB}

The flux of energy injected by DM annihilation, from the
recombination epoch until today through the formation history of DM
halos, results in ionization and heating of the intergalactic
medium.  The ionization and heating can be produced both by the
highly energetic `prompt' photons directly emitted in the
annihilation of two DM particles, and by the lower energy photons
produced by inverse Compton scattering. The latter turns out to be by
far the most important process; in fact, the cross section for
$\gamma e^-$ scattering decreases rapidly with the energy of the
impinging photon, so that low energy photons are more efficient in
removing the electrons from the atoms.  These `primary
reionization' electrons then deposit their energy in the
intergalactic medium via several other interactions, freeing many
more electrons and also augmenting the temperature of the gas. One
way to constrain DM annihilation properties is therefore to look at
the modifications of the CMB spectrum produced by the fact that the
CMB photons meet an opaque medium (with free electrons produced by
the ionization) in their journey from the surface of last scattering.
This is in particular encoded in the total optical depth of the Universe
$\tau$ parameter. $\tau$ is measured by WMAP to be $\tau = 0.084 \pm
0.016$~\cite{WMAP}, of which about 0.038 is due to the low-redshift
reionization ($z < 6$) produced by stars. A DM-induced optical depth
larger than 0.062 (the 1$\sigma$ upper bound of \cite{WMAP}) is
therefore excluded by these arguments.

\begin{figure*}[t]
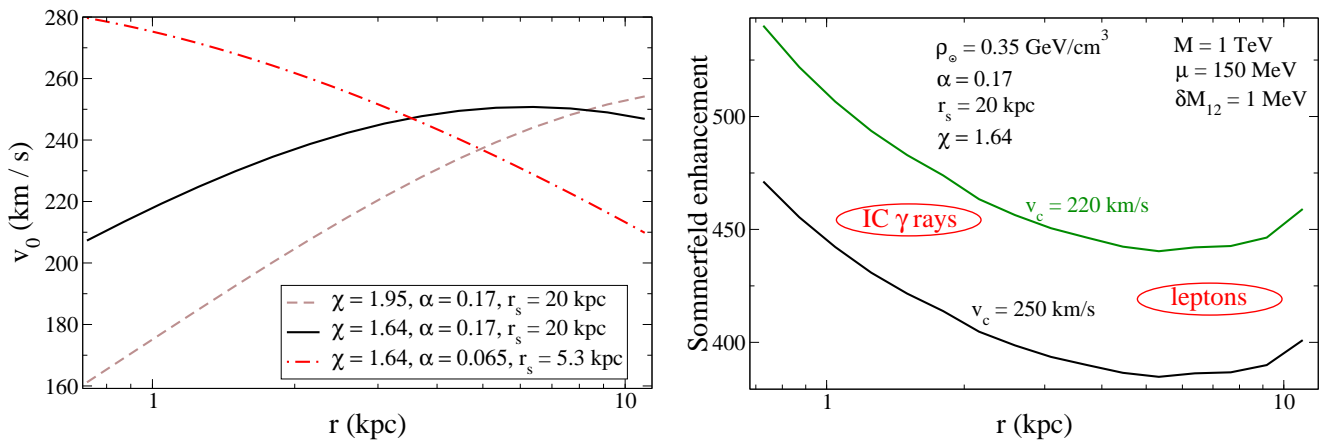

\smallskip \centerline{\epsfxsize=0.48\textwidth\epsfbox{vel-log2.eps} \quad
\epsfxsize=0.47\textwidth\epsfbox{boost-r-log.eps}}
\caption{(a) Left: circular velocity (\ref{bert}) as a function of $r$
for exponents $\chi=1.64$ (includes effects of baryons) and
$\chi=1.95$ (pure dark matter) at standard Einasto parameters
$\alpha = 0.17$ and $r_s = 20$ kpc, and for a 
very cuspy profile (dot-dashed curve).  
(b) Right: velocity-averaged
Sommerfeld enhancement as a function of $r$ for two different
values of the circular velocity $v_c \equiv v_0(r_\odot)$.
The ellipses denote regions in $r$ where the dominantly observed IC $\gamma$ rays
and respectively leptons are produced.}
\label{v0r}
\end{figure*}

This bound has been studied in the literature in several references
  \cite{galli,SPF,hutsiCMB,CIP,kanzakiCMB}, with mutually consistent results.
The fit to the numerical result in \cite{CIP} is
\beq
	\log\sigma_{\rm CMB} v =  -26.3 + 1.15 \log M
\label{cmb_bound}
\eeq
where $\sigma v$ is in cm$^3$/s and $M$ is in GeV. While this was
originally computed for 2-lepton final states, it applies equally
well for the 4-lepton ones in which we are interested. All that
matters is the total amount of energy that is injected in the
primordial intergalactic gas in the form of ``electromagnetically
coupled" final products (electrons, positrons, photons, possibly
hadrons, but not neutrinos). All such final products end up causing
reionization and thus producing free electrons, regardless of the
precise way in which the annihilation occurs. 
 We neglect minor differences possibly introduced by
the difference in the shape of the spectra. The dotted lines in fig.\
\ref{allowed} and \ref{triplet-allowed} show these bounds.
What one sees is that, for the $4e$ case in which we are interested, they rule out the large DM mass portion of the PAMELA-only fit region, but not the PAMELA+Fermi one.

In the present work, we consider the case where the annihilating DM
constitutes a fraction $1/f$ of the total DM.  Since the annihilation
rate scales like $\rho^2\sigma$, the bound (\ref{cmb_bound}) will be
weakened by $\sigma_{\rm CMB}\to f^2\sigma_{\rm CMB}$.

\subsection{Extragalactic gamma ray constraints}

Finally, we briefly review the extragalactic gamma ray
constraints on DM annilation. These refer to the fluxes of prompt
and inverse Compton gamma rays that are produced by annihilations in
all DM halos outside of our own galaxy and throughout the history of
cosmological structure formation. They reach us (properly redshifted)
uniformly from all directions, so that they have to be compared with
the Fermi measurements of isotropic diffuse gamma rays
\cite{fermiisotropic}. A number of references have gone through this
analysis \cite{Profumo,Belikov:2009cx,hutsi,zaharijas}, showing that
the bounds depend very significantly on assumptions about the
parameters of the history of structure formation. {\it E.g.,} for
conservative choices of the halo concentration parameter function,
all of the parameter space in which we are interested is allowed
by these bounds.

\section{Reconciling anomalous leptons with $\gamma$-ray constraints}
\label{results}

We have now discussed most of the necessary ingredients for scanning
over the parameters of the theoretical models and of the DM density
and velocity distributions, to see if there exist any choices that
are consistent with all the observational requirements.  It remains
to explain some details of our methodology for predicting the
effective annihilation cross sections,  both for the lepton signal
and for  the IC $\gamma$ ray signal.  These can differ from each
other
somewhat because of the fact that most of the observed leptons are
produced within 1 kpc of the Sun, whereas the most important IC
$\gamma$ rays come from closer to the galactic center, and the
Sommerfeld enhancement of the cross section is position dependent. 
We explain these details  in the present section, and use them to
obtain the final results.  

\subsection{Position dependence of enhancement factor}
\label{boostr}

As explained above eq.\ (\ref{vavg}), the Sommerfeld enhancement
must be averaged over the phase space of the DM.  Previously
we only considered this in the vicinity of the Sun, but now we must
take into account that the average DM velocity as well as the escape
velocity depend upon $r$.  This dependence has been measured in 
$N$-body simulations, and is shown to follow a scaling relation
predicted by Bertschinger \cite{bert},
\beq
	v_0(r)^3 \propto r^\chi\,\rho(r)
\label{bert}
\eeq
where the exponent $\chi \cong 1.9-2.0$ for pure DM simulations
\cite{Navarro2008},
and it takes lower values $\chi \cong 1.60-1.67$ in simulations
including baryons \cite{tissera}.  The $r$-dependence of the
escape velocity is estimated as in eq.\ (\ref{vesc}).
We plot the $r$-dependence of $v_0$ for the standard $\alpha=0.17$,
$r_s=20$ kpc profile, normalized so that $v_0=250$ at $r=r_\odot$
(the latter choice does not effect the shape of $r$-dependence) in 
figure \ref{v0r}(a).  The shape is strongly dependent upon the choice
of $\chi$, with the larger $\chi$ values giving rise to more
pronounced dependence of $v_0$ upon $r$.

Our choice of $v_0(r)$ differs from one adopted by other authors
\cite{mixdm}, $v_0(r)\sim r^{-1/4}$, that rises as $r\to 0$.   The
$r^{-1/4}$ ansatz was inferred from looking at fig.\ 2 of
ref.\ \cite{Romano}, one of the earlier studies of the effects of
baryons on the halo properties.  (It should be noticed however that
even in this figure, the logarithmic slope of the $z=0$ velocity
profile does not remain constant as $r\to 0$, but starts to turn
downward as in our fig.\ \ref{v0r}(a).)   Based on the more detailed
study \cite{tissera}, we infer that the apparent continued rise of
$v_0$ toward $r=0$ is consistent with our choice (\ref{bert}) if one
also adopts the cuspy profiles found by the simulations that include
baryons.   To illustrate this, we plot an example in fig.\
\ref{v0r}(a)  (dot-dashed curve) using the steepest profile found in
the $N$-body simulation with baryons of ref.\ \cite{tissera}, with
$\alpha = 0.065$ and $r_s = 5.3$ kpc.  The upshot is that
eq.\ (\ref{bert}) predicts that $v_0(r)$ will always reach a maximum at
some $r=r_{\rm max}$ and thereafter fall off as $r\to 0$, but the
value of $r_{\rm max}$ is smaller for cuspier halos.  In the present
work, we will find that cuspy halos are not consistent with
satisfying the inverse Compton constraint, so the 
$v_0(r)\sim r^{-1/4}$ would not be the appropriate one for us to use.

The $r$-dependence of $v_0$ gives rise to  $r$-dependence in the 
velocity-averaged enhancement factor $\langle 
S\rangle$ eq.~(\ref{vavg}) \cite{robertson}.
We illustrate for two different values of the circular velocity at
the solar radius, $v_c = v_0(r_\odot) = 220$ and 250 km/s, adopting
the exponent $\chi =1.64$ which is in the middle of the range for DM
simulations including baryons.  The enhancement factor is stronger near the
galactic center than at the Sun,  implying that IC $\gamma$ rays are
produced more copiously relative to leptons  than would be the case
for a spatially constant boost factor.   This is illustrated in 
figure \ref{v0r}(b) where $\langle S(r)\rangle$ is plotted.  As a
result, the IC bound constrains the models more strongly than if one
ignored this effect.  Like for $v_0(r)$, the shape of  $\langle
S(r)\rangle$ is nearly independent of the value of $v_c$, but depends
strongly on the choice of exponent $\chi$.  Had we chosen the pure DM
value $\chi=1.9$, we would get a much stronger ratio of $\langle
S\rangle$ in the galactic center versus solar regions, which would
make the IC
constraint even more difficult to satisfy.

\begin{figure*}[t]
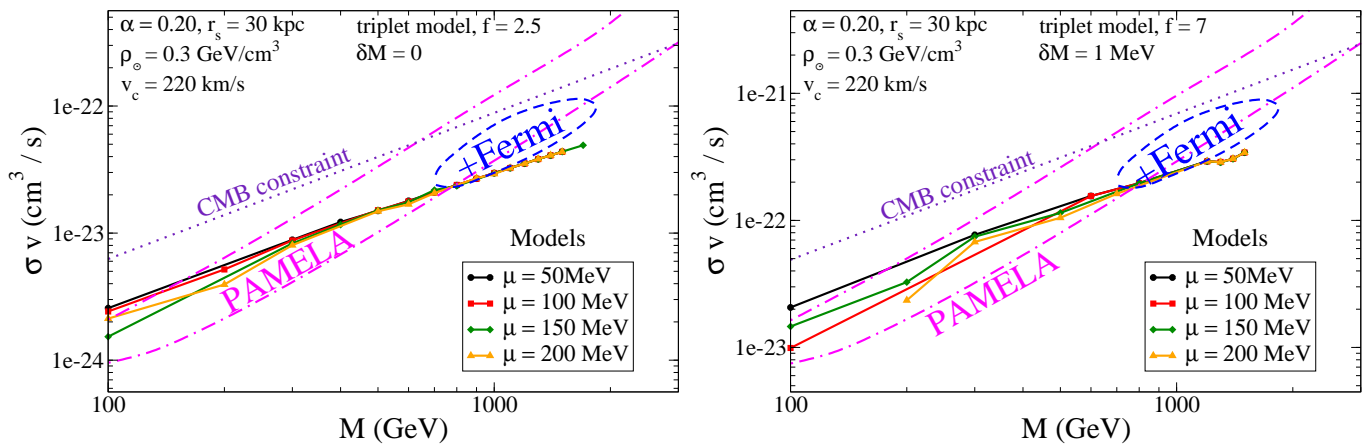

\smallskip 
\centerline{
\epsfxsize=0.5\textwidth\epsfbox{fermi-a20-f25.eps}
\epsfxsize=0.5\textwidth\epsfbox{fermi-a20-f7.eps}}
\caption{Predicted and allowed values of $\sigma v$ versus
$M$ for triplet DM with gauge boson mass $\mu = $ 50, 100, 150 or 200
MeV, and halo parameters $\alpha = 0.20$, $r_s = 30$ kpc,  
$\rho_\odot = 0.3$ GeV/cm$^3$.  Only predicted values consistent with
IC constraint are shown.  Left: mass splitting $\delta M = 0$, DM 
fraction $1/f = 0.4$; right $\delta M = 1$ MeV, $1/f=1/7$.}
\label{triplet-fits}
\end{figure*}

\begin{figure*}[t]
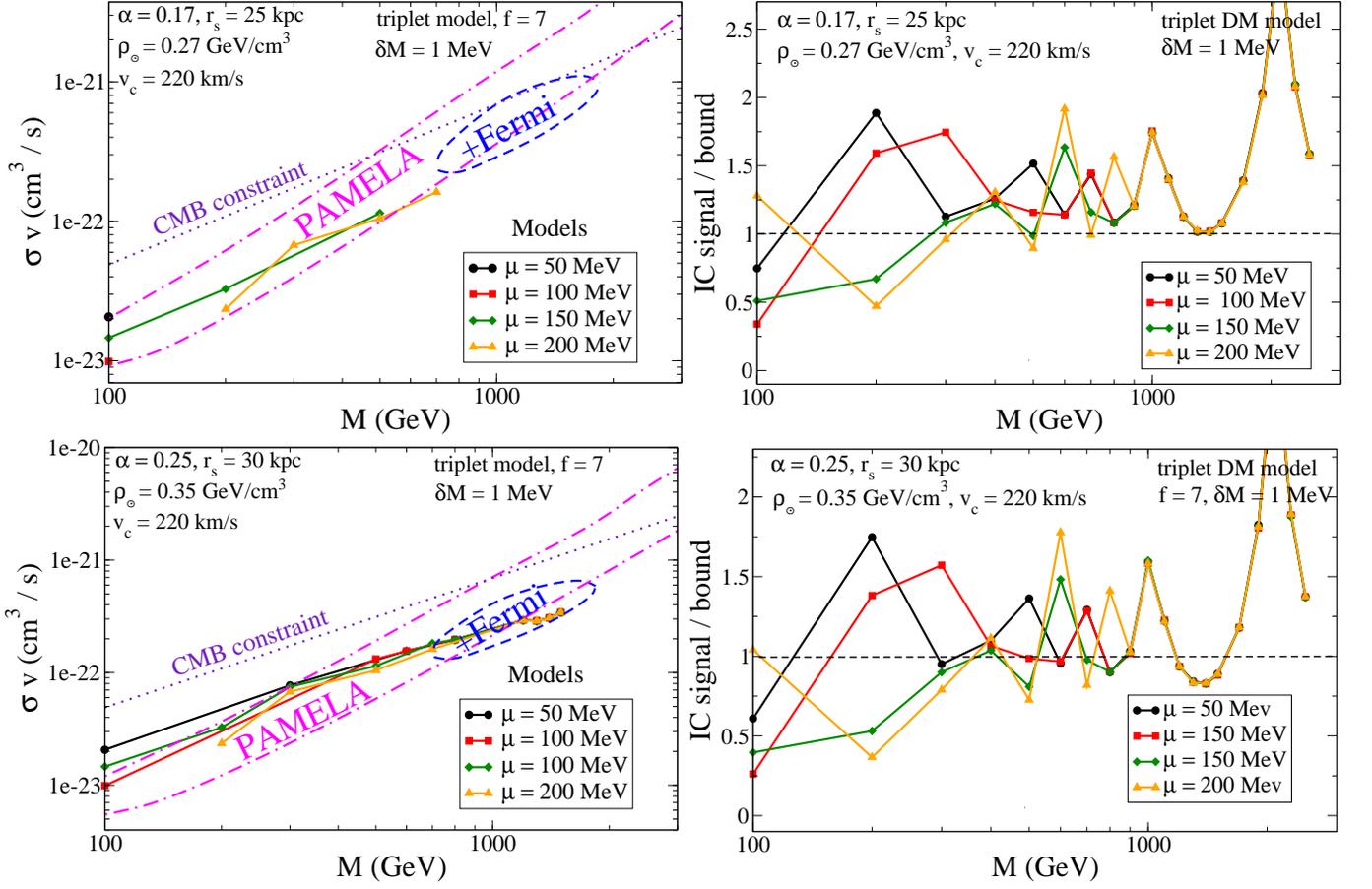

\smallskip 
\centerline{
\epsfxsize=0.51\textwidth\epsfbox{fermi-a17-f7.eps}
\epsfxsize=0.49\textwidth\epsfbox{bound-ratio-a17-f7.eps}}
\centerline{\epsfxsize=0.51\textwidth\epsfbox{fermi-a25-f7.eps}
\epsfxsize=0.49\textwidth\epsfbox{bound-ratio-a25-f7.eps}}
\caption{Left: Predicted and allowed values of $\sigma v$ versus
$M$ for triplet DM with gauge boson mass $\mu = $ 50, 100, 150 or 200
 MeV, $\delta M = 1$ MeV, and DM density fraction $1/f = 1/7$.  
Only the model predictions that are consistent with the IC
constraint are shown.
Right: ratio of predicted versus maximum allowed IC $\gamma$ ray
cross section for same models.  Top row: $\alpha = 0.17$, 
$r_s= 25$ kpc, and
$\rho_\odot = 0.27$ GeV/cm$^3$; bottom row:
$\alpha = 0.25$, $r_s = 30$ kpc  
and $\rho_\odot = 0.35$ GeV/cm$^3$.}
\label{triplet-allowed}
\end{figure*}

\subsection{Estimation of lepton and IC $\gamma$-ray signals}

Our procedure now is to define the cross section for production of
$e^\pm$ (and below, that of IC $\gamma$ rays), using the partial cross section
(\ref{sigma_part}) and enhancement factors that have been averaged over the
appropriate regions of space.  The average is done
with a weighting factor of $r^2\rho^2$,
appropriate for annihilations occuring in a shell of radius $r$ and 
thickness $dr$ centered on the sun.  Assuming that the observed leptons 
have diffused no further than $\Delta r = 1$ kpc \cite{ediff}, the leptonic enhancement
factor is
\beq
	B_{e^\pm} =   {\int_{r_\odot -\Delta r}^{r_\odot +\Delta r} dr\, r^2\,  
\rho^2(r)\, \langle S(r)\rangle \over
\int_{r_\odot -\Delta r}^{r_\odot +\Delta r} dr\, r^2\, \rho^2(r)} 
\label{Bepm}
\eeq
and the leptonic cross section is $\sigma_{e^\pm} = b_{11} B_{e^\pm} 
\sigma_{0} f$, where $\sigma_{0}$ is the reference cross section that
would give the correct relic density, $\sigma_0 v = 3\times 10^{-26}$
cm$^3$/s.  The resulting $\sigma_{e^\pm}v$ is then
compared to the PAMELA and Fermi allowed regions from \cite{papucci},
rescaled by $(\rho_{\odot,0}/\rho_\odot)^2 f^2$ if necessary.  (We do not
apply a correction factor for changing $\alpha$ or $r_s$
as in the case of the IC constraint, because  the behavior
of $\rho(r)$ in the vicinity of $r_\odot$ does not change
significantly as a function of $\alpha$ or $r_s$.)
This tells us if the predicted leptonic signal is sufficiently close to the
observed one.

\begin{figure*}[t]
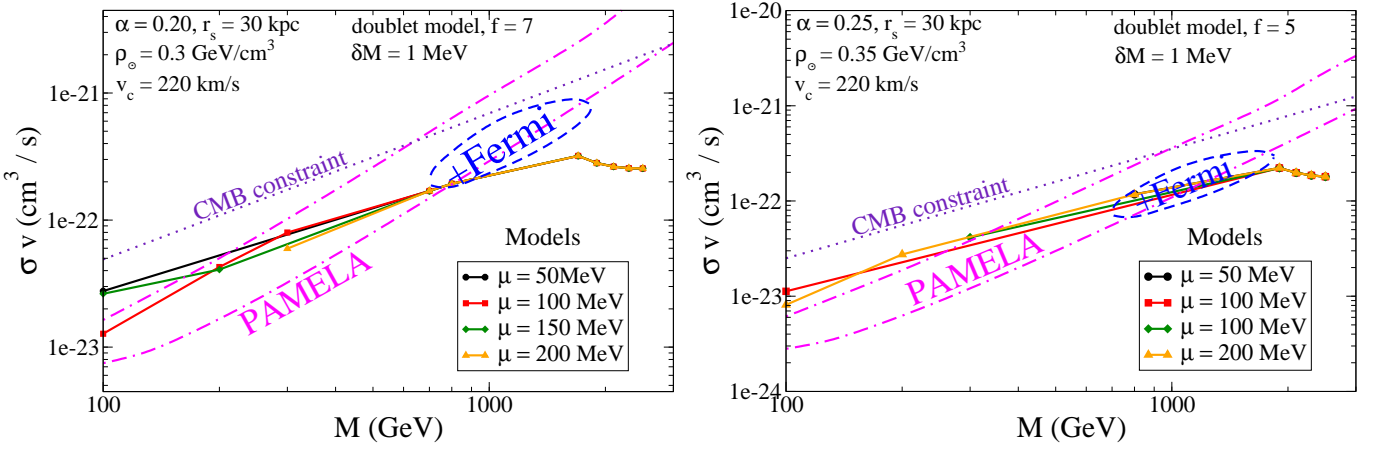

\smallskip 
\centerline{
\epsfxsize=0.5\textwidth\epsfbox{doublet-a20-f7.eps}
\epsfxsize=0.5\textwidth\epsfbox{doublet-a25-f5.eps}}
\caption{Same as left panels of fig.\ \ref{triplet-allowed}, but
for the doublet DM model.  Left: 
$\alpha = 0.20$, $r_s = 30$ kpc,  
$\rho_\odot = 0.3$ GeV/cm$^3$, $f=7$; Right:
$\alpha = 0.25$, $r_s = 30$ kpc, 
$\rho_\odot = 0.35$ GeV/cm$^3$, $f=5$.}
\label{doublet-allowed}
\end{figure*}

For the IC $\gamma$ ray signal we follow a similar approach, but now
$\langle S(r)\rangle$ should be averaged over a radial interval that
corresponds to the path followed by the relevant $\gamma$ rays.   
Since the IC signal arrives along a line of sight rather than from a 
fixed volume, the weighting factor is $\rho^2$ rather than
$r^2\rho^2$.  We thus take 
\beq
	B_{\gamma} =   {\int_{r_i}^{r_\odot} dr\,  
\rho^2(r)\, \langle S(r)\rangle \over
\int_{r_i}^{r_\odot} dr\,  \rho^2(r)} 
\label{Bgamma}
\eeq
where 
$r_{i} = \tan(5^\circ)\, r_\odot = 0.73$ kpc is a cutoff 
due to the fact that the
skymap in fig.\ \ref{skymap} does not use the inner 5$^\circ$
for setting the IC bounds.
The predicted $\gamma$ ray signal corresponds to the cross section
$\sigma_{\gamma} = b_{11} B_{\gamma} \sigma_{0} f$.  This should be
below the bound 
\beq
	\sigma_{\gamma}  < \sigma_{4e}\left({\rho_{\odot,0}\over
	\rho_\odot}\right)^2 
	\left({\bar J_0\over \bar J}\right)\, f^2
\label{ICbound}
\eeq
implied by eqs.\ (\ref{bounds},\ref{s4es},\ref{approx}).

\subsection{Search of parameter space}

We have surveyed the predictions of the models for a range of DM halo
parameters $\rho_\odot$, $\alpha$, $r_s$, $v_c$ in search of examples
that can satisfy all the constraints.  There are two principal
challenges to confront.  First, as we illustrated in fig.\
\ref{boost}, the Sommerfeld enhancement tends to be too large to fit
the lepton signals,  especially if the mass splitting is nonzero. 
Second, the IC constraint rules out large boost factors unless the DM
halo parameters are taken to be noncuspy, {\it i.e.,} large values of
$\alpha$ and $r_s$.  On the other hand, $N$-body simulations favor
smaller values of $\alpha\cong 0.16-0.17$ for galaxies of the size of
the Milky Way \cite{Navarro2008,gao,ghalo}. 

The upshot is that a compromise must be made: to avoid extremely
large values of $\alpha$, one needs to marginally satisfy the IC
constraint by making the annihilation rate as small as possible,
which also pushes the lepton signal to the lower boundary of its
allowed region.  In order to achieve this reduction in the
annihilation rate, we are forced to assume that the annihilating DM
only constitutes the fraction $1/f$ of the total mass density, by
increasing the gauge coupling as in (\ref{relic_density_f}) to
suppress the relic density.  Since the Sommerfeld enhancement rises
as a function of the mass splitting, there is a trade-off between the
parameters $f$ and  $\delta M$.  

We illustrate the best-case scenarios in figure \ref{triplet-fits}
for the triplet DM  model with $\delta M=0$ and $\delta M=1$ MeV,
which shows the model predictions for $\sigma_{e^\pm} v$ and the
$3\sigma$ confidence level allowed values for PAMELA/Fermi for the
two cases. In the first case a sufficiently small annihilation rate
is found using the maximum DM fraction $1/f = 0.4$, while in the
second, it must be reduced to the level of $1/7$.  These examples
illustrate the optimal compromise for satisfying the IC constraint;
we are able to just accommodate it using the large, noncuspy Einasto
parameter choices  $\alpha = 0.20$,  $r_s=30$ kpc.\footnote{Ref.\
\cite{Lin} also finds $\alpha = 0.20$ in a recent fit to the data
including $\gamma$ rays.}  The lepton rate is barely consistent with
the PAMELA/Fermi allowed region for a DM mass of 800 TeV.  The
relatively low  circular velocity $v_c = 220$ km/s we have adopted
also helps to  satisfy the IC bound, since as figure \ref{v0r}
implies, the ratio $B_\gamma/B_{e^+}$ is somewhat smaller for low
$v_c$ than for high $v_c$.

To highlight the sensitivity of the IC constraint and lepton signal
to the cuspiness of the halo profile, we present results of our
search using the neighboring Einasto parameters  $\alpha=0.17$, 0.25 
and the triplet DM model, in figure \ref{triplet-allowed}.  The 
left-hand panels are similar to fig.\ \ref{triplet-fits}, while  the
right hand ones plot the ratio of $\sigma_{\gamma}$ to the IC bound
(\ref{ICbound}).   The points above the dashed line in the right-hand
panels are ruled out by the IC bound, and the figures show how this
becomes more severe for smaller $\alpha$.
 The left panels  also  show the CMB constraint
from reionization of the early universe; this is satisfied by a much
wider margin than is the IC constraint, as already anticipated. 
Figure \ref{triplet-allowed} illustrates that for the usually
preferred choice $\alpha = 0.17$, only compatibility with PAMELA can
be  achieved, but not simultaneously with the Fermi lepton anomaly.
Only models with  $M \lsim 700$ GeV are consistent with the IC
constraint, whose strength is an increasing function of $M$.  

The tension we find here is not ameliorated by considering the
doublet DM model instead of the triplet.  Given that the gauge
couplings $\alpha_g$ and cross-section ratios $b_{11}$ are rather
different between the two models, their predictions are suprisingly
similar. (This is partly understandable in that we adjust the
tree-level cross sections of the two models to be the same, but the
Sommerfeld enhancement introduces extra dependence upon $\alpha_g$,
so there seems to be an accidental cancellation of the effect of
$b_{11}$ by the increase in the Sommerfeld boost.)   Moreover the
dependence upon the halo parameters is very similar between the two
models.  In fig.\ \ref{doublet-allowed} we show the predictions and
constraints on $\sigma v$ versus $M$ for the doublet model, using the
$\delta M = 1$ MeV mass splitting, and two of the same sets of halo
parameters (the less cuspy ones) as we considered for the triplet
model.  The results are difficult to distinguish from those of the
triplet model.

\section{Discussion}
\label{discussion}

We have tried to carefully assess the likelihood that some of the
best 
theoretically motivated models of annihilating multistate dark
matter can explain the PAMELA/Fermi/H.E.S.S.\ lepton excesses,
considering important details of the models whose implications could be
missed by a more generic, model-independent investigation.   As well,
we take into account simulations and observations of the DM halo
properties, which have a crucial impact on the viability of the
scenario.  The main challenge to the models is the difficulty of
satisfying constraints from Fermi due to the production of inverse
Compton (IC) $\gamma$ rays produced by the high-energy leptons
scattering on galactic radiation.

Among the most important model-dependent features, which affect the
ease of  satisfying the constraints, are that each annihilation
produces two dark gauge bosons, hence four leptons, and these are all
$e^\pm$ as opposed to heavier charged particles if the gauge boson
masses are lighter than $\sim 210$ MeV.  The IC constraints for this
case are less severe than for models that produce only two leptons,
or ones that produce $\mu^\pm$ in addition to $e^\pm$.   Moreover the
gauge coupling is bounded from below by the requirement that the
relic density not exceed what is observed.  We find that the boost
factor from Sommerfeld enhancement is too large to satisfy the
constraints, unless the density of the annihilating DM is suppressed
by a factor $1/f < 0.4$; hence one would need an additional component
of nonleptophilic DM to make up the rest.  Using recent results for
the Sommerfeld enhancement factor of multistate DM \cite{slatyer},
this problem is exacerbated when the mass splitting is nonzero; one
needs $1/f \sim 0.14-0.2$ if $\delta M \sim 1$ MeV. 

A main result is that the Einasto parameters which determine the
shape of the DM halo must take large values $\alpha \gsim 0.20$ and
$r_s \sim 30$ kpc in order to barely satisfy the IC constraint and
have marginal consistency with the observed lepton excesses.   In the
context of pure dark matter simulations, such a large value of
$\alpha$ is above the norm for a galaxy of the size of the 
Milky Way.   Ref.\  \cite{gao} finds a correlation between $\alpha$ and
the virial mass of galaxies $M_{\rm vir}$, $\alpha = 0.155 + 0.0095\,
\nu^2$, where  $\nu = \delta_{\rm crit}(z)/\sigma(M_{\rm vir})$ are
functions arising in the Press-Schechter formalism
\cite{press-schechter} such that  $\delta_{\rm crit} = 1.686$ at
redshift $z=0$, and $\sigma > 2$  \cite{loeb} for a galaxy such as
ours with $M_{\rm vir} = 10^{12} M_\odot$ \cite{xue}.  Thus $\nu <
0.843$, which implies  $\alpha < 0.162$.  However this is only true
for the mean value, and a fluctuation as large as $\alpha=0.20$ for a
galaxy like ours is possible \cite{navarro-com}.  On the other hand,  recent results
incorporating baryons indicate that $\alpha$ is decreased relative to
its pure DM value due to adiabatic contraction, which exacerbates the
conflict.

\medskip

We conclude that there is a distinct tension for  annihilating models
to consistently account for all the lepton excesses, while it is
easier to explain that of PAMELA alone using smaller DM masses
$M<400$ GeV.   A possible loophole that we
have not considered here is the suggestion that leptons originating
from subhalos of our main halo could provide a substantial fraction
of the observed  leptonic excess \cite{subhalo}.  If this is the
case, the production  of IC $\gamma$ rays from the galactic center
would be reduced, providing another way to weaken the constraint. 
How significant this effect could be is under investigation.

\bigskip

{{\bf Acknowledgments}. We thank Fang Chen, Ilias Cholis, Julio Navarro, 
Pasquale Serpico, Tracy Slatyer and Patricia Tissera for helpful
discussions and correspondence. We are especially grateful to Alessandro
Strumia for detailed information pertaining to ref.\ \cite{papucci}.
M.C.\ thanks the EU Marie Curie Research \& Training
network ``UniverseNet" (MRTN-CT-2006-035863) for support.  J.C.\ is
supported by the Natural Sciences and Engineering Research Council
(Canada).}

\appendix

\section{Cross section ratios for $\chi_1\chi_1\to e^\pm$}
\label{brfr}

In this appendix we compute the ratio of the cross section for
ground state DM annihilation into $4e$ final states relative to the 
total annihilation cross section in the early universe.  

\subsection{Triplet model}

The total cross section for $\chi\chi\to B B$ annihilation was
derived for DM in any representation of SU(2) in appendix C of ref.\
\cite{nonabelian}.  For the triplet model, the averaged squared
matrix element can be expressed in the form
$\frac13|{\cal M}_{11}|^2 + \frac23 |{\cal M}_{12}|^2$,
where $|{\cal M}_{ij}|^2$ is the matrix element for $\chi_i\chi_j$,
summed over all possible final states.  The factor of $\frac13$ comes
from averaging over the colors of $\chi_j$.   On the other hand, the
matrix element for $\chi_1\chi_1\to B_2 B_2$ was shown to be just
half of $|{\cal M}_{11}|^2$, since the latter includes 
$\chi_1\chi_1\to B_3 B_3$.  Therefore the ratio $b_{11}$ for the
single process $\chi_1\chi_1\to B_2 B_2$ is given by
\beq
	b_{11} = {\frac12 |{\cal M}_{11}|^2\over 
	\frac13|{\cal M}_{11}|^2 + \frac23 |{\cal M}_{12}|^2}
	= {\frac12\cdot 8 \over \frac13(8 + 5/2)} = \frac87
\eeq
using the results of \cite{nonabelian} for the matrix elements.
In units of $g^4$, $|{\cal M}_{11}|^2 = 2(|t|^2 + |u|^2)$ where 
$|t|^2$ and $|u|^2$ stand for the contributions from 
the $t$ and $u$ channels, respectively, and take the values 
$|t|^2 = |u|^2 = 2$.  Likewise $|{\cal M}_{12}|^2 = |t|^2 + |s|^2
+ |st|$ with $|s|^2 = -19/4$ and $|st| = 4$.  The inteference
between $t$ and $u$ channels vanishes, $|tu|=0$.  

\subsection{Doublet model}

For the doublet model, the matrix element for 
 $\chi_i\chi_j\to B_a B_b$ can be expressed in the form
\beq
	{\cal M}_{ij}^{ab} = \frac14 \tau^b_{jk}\tau^a_{ki} t
	+ \frac14 \tau^a_{jk}\tau^b_{ki} u + 
	\frac{i}{2}\tau^c_{ji}\epsilon^{abc} s
\eeq
where $s,t,u$ are the spinor parts of the amplitude in the respective
channel.  The factors of $1/2$ are for the normalization of the SU(2)
generators.  To find the averaged cross section for the early
universe, we sum over $a,b$ and average over $i,j$.  This gives
\beqa
	\langle |{\cal M}|^2\rangle &=& 
	\frac14\left[\frac{18}{16}\left(|t|^2 +|u|^2\right) + 
	\frac{12}{4}|s|^2 + \frac{12}{8}\left(|st| + |su|\right)\right]
	\nonumber\\ &=& \frac{9}{16}
\eeqa
where we used the values given above for $|t|^2$, $|u|^2$, $|s|^2$,
$|st|$ and also $|su|=|st|$.  This expression agrees with the general
result $\langle |{\cal M}|^2\rangle =
\frac{3j(j+1)}{2j+1}[j(j+1)-1/4]$ for the spin-$j$ representation of 
SU(2) found in \cite{nonabelian}, when $j=1/2$.  We need to compare
this result with that of the exclusive channel $\chi_1\chi_1\to B_2
B_2$, 
\beq
 |{\cal M}_{11}^{22}|^2 = \frac{1}{16}\left(|t|^2 +|u|^2\right) = 
	\frac14
\eeq
The ratio of these would give $(\frac14)/(\frac{9}{16}) = 4/9$.  
However unlike the triplet model, there is an additional process that
can give the $B_2$ gauge boson, namely $\chi_1\chi_1\to B_1 B_2$,
which has
\beqa
	|{\cal M}_{11}^{12}|^2 &=& \frac{1}{16}\left(|t|^2 +|u|^2\right)
+ \frac14|s|^2 + \frac18\left(|st| + |su|\right)\nonumber\\
	& =& \frac{1}{16}
\eeqa
This should be doubled to take into account the 
$\chi_1\chi_1\to B_2 B_1$ contribution, but we halve it again to
account for the fact that only one $B_2$ gauge boson is produced, so
it gives half as many leptons as the $B_2 B_2$ final state.
Hence we find that
\beq
	b_{11} = {\frac14+\frac1{16}\over \frac{9}{16}} = \frac59
\eeq

\section{Multistate Sommerfeld enhancement}
\label{bfa}
The Sommerfeld enhancement factor for multicomponent 
DM annihilation is different from the that of the single component 
case.  In particular if the gauge couplings are off-diagonal,
$\bar\chi_1\slashed{B}\chi_2$ and if there is a mass difference
$\delta M$ between $\chi_1$ and $\chi_2$, ref.\ \cite{slatyer}
finds that the enhancement factor is given by
\begin{widetext}
\beq S = \frac{2 \pi }{\epsilon_v}\sinh\left(\frac{\epsilon_v \pi
}{\nu}\right) \left\{ \begin{array}{cc}
\frac{1}{\cosh\left(\epsilon_v
\pi/\nu\right)-\cos\left(\sqrt{\epsilon_\delta^2-\epsilon_v^2} \pi
/\nu+2 \theta_-\right)} & \quad \epsilon_v < \epsilon_\delta, \\ \\
\frac{\cosh\left(\left(\epsilon_v+\sqrt{-\epsilon_\delta^2+\epsilon_v^2}\right)
\pi/2 \nu\right)
\text{sech}\left(\left(\epsilon_v-\sqrt{-\epsilon_\delta^2+\epsilon_v^2}\right)
\pi/2 \nu \right)}{
\cosh\left(\left(\epsilon_v+\sqrt{-\epsilon_\delta^2+\epsilon_v^2}\right)
\pi /\nu\right)-\cos(2 \theta_-)} &\quad \epsilon_v >
\epsilon_\delta. \end{array} \right. \label{slat}\eeq 
\end{widetext}
with the definitions $\epsilon_v = v/(c\alpha_g)$ ($v$ being the
center-of-mass velocity of one of the DM particles),
$\epsilon_\phi = \mu/(M\alpha)$, and 
$\nu = \frac12\epsilon_\phi(1 + \sqrt{1 + 4/(\epsilon_\phi r_M)})$.
$r_M$ is a radial coordinate value defined such that $V(r_M) =$
max$(\epsilon_\delta^2/2,\epsilon_\phi^2)$, where
$\epsilon_\delta = \sqrt{2\delta M/ M}/\alpha_g$ and $V(r) =
e^{-\epsilon_\phi r}/r$.  $\theta_-$ is a function defined as an
integral that must be done numerically (see eq.\ (4.8) and footnote
3 of \cite{slatyer}).

\section{Escape velocity}
\label{escvel}

In general, the escape velocity from a given radius is given in terms
of the gravitational potential
\beq
	\frac12 v_{\rm esc}^2 = -\Phi(r) = G\int_r^\infty 
	{dR\over R^2}
	M(R)
\eeq
where $M(r)$ is the mass enclosed within radius $r$.  If all of the
matter in the galaxy was dark, we could simply integrate $\rho(r)$
to find $M(r)$, but baryons constitute a significant fraction of the
matter in the inner part of the galaxy, so this would give an
underestimate of $v_{\rm esc}$.  Instead, one can try to infer the
shape of $M(r)$ from rotation curves, which measure the circular
velocity $v_c(r)^2 = G {M(r)/r}$.  We thus have the relation
\beq
	v_{\rm esc}^2 = 2\int_r^\infty {dr\over r} v_c^2(r)
\label{vesc_est}
\eeq

For flat rotation curves, $M(r)$ increases linearly with $r$, and
$v_{\rm esc}^2$ goes like $-\ln(r)$.  Recent observations of the
Milky Way \cite{xue} indicate that its circular velocity is nearly 
constant to a radius of $r_{10} \equiv 10$ kpc, and falls slowly like
$r^{p-1}$ between $r_{10}$ and 60 kpc, where $p<1$.  This corresponds
to $M(r) \sim r^{2p-1}$ in the region between 10 and 60 kpc.  Since $v_c$
falls from 200 km/s at $r_{10}$ to 170 km/s at 60 kpc, we infer that
$p = 0.856$.  This tells us that $v_c = 220$ km/s for $r < r_{10}$
and $v_c = (220$ km/s)$\times (r/r_{10})^{p-1}$ in the region between
$r_{10}$ and 60 kpc.  

We still need an estimate of how $v_c$ behaves at $r > 60$ kpc in
order to do the integral (\ref{vesc_est}).  The most radical
assumption would be that there is no more mass beyond this radius, so
that $v_c$ drops like $1/\sqrt{r}$ beyond this point.  However, the
appearance of the rotation curve at the highest measured radii does
not suggest such a change; moreover we have an estimate of the total
mass of the galaxy through its virial mass $M_{\rm vir}$, determined
to be $10^{12} M_\odot$ in ref.\ \cite{xue}.  Therefore it would seem
reasonable to assume that $M(r)$ continues to grow with the same
power law $r^{2p-1}$ out to a radius that contains $M_{\rm vir}$, after
which it stops growing.  In the present case that radius corresponds
to $r_{\rm vir} = 175$ kpc.   We therefore take $r_{\rm vir}$ as our
estimate of where $M(r)$ ceases to grow.  In the region $r< r_{10}$,
this gives 
\beqa
	v_{\rm esc}^2 &=& 2 v_c^2 \Bigg[	
	\ln\left({r_{10}\over r}\right) \nonumber\\
&+& \frac{1}{2(1-p)}
	\left( 1 + (1-2p) \left({r_{10}\over r_{\rm
vir	}}\right)^{2(1-p)}
	\right) \Bigg]
\eeqa

\end{document}